\documentclass[aps,pre,twocolumn,amssymb,amsmath,showpacs,amsmath,amssymb]{revtex4}
\usepackage{latexsym}
\usepackage[dvips]{graphicx}
\usepackage{dcolumn}
\usepackage{bm}
\usepackage{url}

\begin{document}

\title{Irreversibility in a simple reversible model}

\author{Juraj \surname{Kumi\v{c}\'{a}k}}
\email{Juraj.Kumicak@tuke.sk}
\affiliation{Department of Thermodynamics, Technical University,\\
Vysoko\v{s}kolsk\'{a} 4, 042 00 Ko\v{s}ice, Slovakia\\
(Received 13 May 2004; published 12 January 2005)}
\begin{abstract}
This paper studies a parametrized family of familiar generalized baker maps, viewed as simple models of time-reversible evolution. Mapping the unit square onto itself, the maps are partly contracting and partly expanding, but they preserve the global measure of the definition domain. They possess periodic orbits of any period, and all maps of the set have attractors with well defined structure. The explicit construction of the attractors is described and their structure is studied in detail. There is a precise sense in which one can speak about absolute age of a state, regardless of whether the latter is applied to a single point, a set of points, or a distribution function. One can then view the whole trajectory as a set of past, present and future states. This viewpoint is then applied to show that it is impossible to define \emph{a priori} states with very large ``negative age''. Such states can be defined only \emph{a posteriori}. This gives precise sense to irreversibility --- or the ``arrow of time'' --- in these time-reversible maps, and is suggested as an explanation of the second law of thermodynamics also for some realistic physical systems. 
\end{abstract}


\maketitle

\section{\label{Introduction}Introduction}

This paper is devoted to an analysis of one aspect of the second law of thermodynamics, namely its statement about irreversibility. However, I am not going to \emph{prove} essentially new facts concerning realistic physical systems. What I actually want to do is to \emph{illustrate} the origin of the irreversible behavior of time-reversible systems using an extremely simple model. The reason for such an approach is that, in my view, the second law (at least in its most general formulations) does not formulate new \emph{facts} about dynamical systems, it just describes their properties in a new form --- that is why it is being derived from the underlying dynamics. One could say that the problems usually associated with substantiation of this law are not so much of a physical as of a conceptual nature. In such a context it may be acceptable to rely on demonstrations instead of proofs, since the topic is more a question of semantics and interpretation.

Microscopic laws of molecular dynamics are invariant with respect to the time reversal: only the momenta change their signs upon the transformation $t \to -t$. This implies that if these laws allow some evolution of a system, they allow also an evolution in which the system passes through the same spatial configurations as the original ones but in the reversed order (and with reversed velocities). If the same systems are viewed macroscopically, they evolve, on the contrary, in one direction only: they demonstrate irreversibility which is formulated in the second law. According to the latter, only evolution leading to growth or preservation of disorder is observable, or in other words, there is an arrow of time. This conceptual asymmetry represents a fundamental problem, called the problem of irreversibility --- known for more than a century --- which does not seem to be completely solved up to the present time.

There is a plethora of theories trying to explain the origin of macroscopic irreversibility. The approaches can be roughly subdivided into those treating ensembles (for overview see e.~g.~\cite{Mackey92} and references therein) and the others studying individual systems (see, e.~g.~\cite{Spohn91}).

In isolated systems, the irreversibility is usually being reduced to asymmetry in possibilities to prepare initial states which would evolve to equilibrium as compared to those evolving away from it~\cite{KumicakBrandas87a, Spohn97}. In open systems, the classical approach is to view the environment as a source of random perturbations~\cite{Xavier92, KumicakHemptinne98}, thus actually substituting deterministic systems by stochastic ones. In such a way, however, the most appealing aspect of the problem --- the reconciliation of microscopic reversibility with macroscopic irreversibility --- is being lost.

Recently, a new promising approach to solution of the problem has been undertaken, studying, among other things, the simple model called the rotated baker map~\cite{HooverRev}, defined on the unit square in the following way:

\begin{equation}
B_r(x,y) = \left\{ \begin{array}{ll}
               (2x/3,\ 3y) & \mbox{for $ y < 1/3$} \\
               (x/3+2/3,\ 3y/2-1/2) & \mbox{for other $y$.}
               \end{array}
       \right.
\label{rotated}
\end{equation}
This model generalizes the well-known ``classical'' baker map,

\begin{equation}
C(x,y) = \left\{ \begin{array}{ll}
               (2x,\ y/2) & \mbox{for $ 0 \leq x \leq 1/2$} \\
               (2x - 1,\ y/2 + 1/2) & \mbox{for other $x$,}
               \end{array}
       \right.
\label{classical}
\end{equation}
and has many interesting properties.

In this context, we will be interested in further generalization of Eq.~(\ref{rotated}), which represents a very simple model enabling to demonstrate many features of interrelation between reversibility and irreversibility.

In the following discussion, we will be frequently encountering the notions of measure and dimension, so that I find it meaningful here to give a brief description of what is meant by them.

Avoiding the use of excessively technical language, we can say that \emph{measure} $\mu$ on $\mathbb{R}^n$ is a set function assigning a non-negative number to any subset of $\mathbb{R}^n$ in such a way that the measure of the empty set is zero, the measure of a subset $A \subset B$ is $\mu(A) \leq \mu(B)$, and the measure of the union of subsets is less than or equal to the sum of the measures (the strict equality holding only in the case of disjoint subsets). This notion is a generalization of that of area or volume and it frequently reduces to it.

A specific kind of measure is the \emph{dimension} of a set, which is again a generalization of our intuitive notion concerning physical (topological) dimension. There are many ways to define and calculate it, leading to different values. The standard set of dimensions~\cite{Schuster88} is based on partitioning the phase space into a finite number $N(\epsilon)$ of disjoint $\epsilon$-cells (boxes) and considering the probability $p_i(\epsilon)$ (the so-called natural measure) of finding points of the set in the box $i$. Of course, the probability has to be normalized so that

\begin{displaymath}
\sum_{i=1}^{N(\epsilon)} p_i(\epsilon)=1.
\end{displaymath}
The dimension is then calculated, for any integer $s \geq 0$, according to the formula

\begin{equation}
D_s = \frac{1}{s-1} \lim_{\epsilon \to 0} \frac{\ln I(s,\epsilon)}{\ln \epsilon},
\label{dimension}
\end{equation}
where

\begin{displaymath}
I(s,\epsilon) \equiv \sum_{i=1}^{N(\epsilon)} p_i^s.
\end{displaymath}

Specifically, for $s \to 0$ we obtain the Hausdorff (or box-counting) dimension

\begin{equation}
D_0= - \lim_{\epsilon \to 0} \frac{\ln [N(\epsilon)]}{\ln \epsilon},
\label{Hausdorff}
\end{equation}
and for $s \to 1$ the information dimension

\begin{equation}
D_1=  \lim_{\epsilon \to 0} \frac{\sum_{i=1}^{N(\epsilon)} p_i \ln p_i}{\ln \epsilon},
\label{information}
\end{equation}
which is commonly used to describe basic properties of fractals. In effect, the information dimension of a set of points gives crude information about its inhomogeneity (hence the name).

It may be worth mentioning that for $s \to 2$ we obtain the correlation dimension. One can also prove under rather general conditions that $D_q \leq D_r$ for $q<r$.

\section{\label{definitions}Definitions of reversibility}

To introduce shortly the notion of (ir)reversibility, consider a continuous dynamical system (a flow) defined in a phase space $\Gamma$ and described by differential equations. If $S_t$ denotes the evolution operator, taking the present state $\gamma_0 \in \Gamma$ to the future one, $\gamma_t = S_t \gamma_0$, then the present state can be retraced into the past as well. If this can be achieved by applying a well-defined operator $S_{-t}$ (originating in differential equations), i.~e., $S_{-t} \gamma_0 = \gamma_{-t}$, then reversibility of the dynamics means that $S_{-t}$ is defined. However, the inverse operator $S_{-t}$ may differ significantly from $S_t$, so that the dynamics of the inverse evolution may be different from that of the forward one. If, however, $S_{-t}$ and $S_t$ differ just in the sign of the para\-meter $t$, such reversibility reduces in a functional analytic approach to the statement that the family of operators $\{S_t\}$ represents a group, defined by a generator, the latter being closely related to the differential equations describing the dynamics~\cite{KumicakBrandas87a}.

Another definition is not so much concerned with changing the direction of evolution as it is with the possibility to ``reverse'' the final state $\gamma_t$ to a reversed one $T \gamma_t$ (typically by reversing the directions of all momenta) and with application of $S_t$ to the latter. The dynamics is then said to be time-reversible if
\begin{equation}
S_t T \gamma_t = T \gamma_0 \mbox{ for all }\ t>0.
\label{reversible}
\end{equation}
The question about the existence of $S_{-t}$ is not so important here and therefore such an approach seems to be more general than the previous one.

Obviously, the properties of the reversal operator $T$ will depend on $S_t$ in general. We expect to have different operators $T$ for different evolution operators $S_t$. Therefore, $T$ can be any transformation which transforms a final state of evolution into the initial state for the same evolution --- fulfilling, of course, the property~(\ref{reversible}). Every such transformation will be evidently idempotent, i.~e., $T^2 = I$, and therefore $T^{-1} = T$.
\begin{figure}[hbt]
	\centering
		\includegraphics[width=60mm]{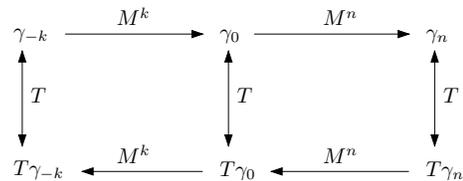}
		\caption{Commutativity as the defining property of reversibility.}
	\label{commute}
\end{figure}

In real physical systems, the operator $T$ has the evident meaning of the change of momenta: $p_i \to -p_i$. The fact that for time-reversible systems we have $S_t T \gamma_t = T \gamma_0$, is just the \emph{consequence} of the physical laws there. If, however, we want to speak about reversibility of systems in which there is no analogy to momenta --- as is, e.~g., the case of two-dimensional maps which will interest us in the following --- we have to accept this consequence as a \emph{definition} and denote as reversible all maps $M$, for which there exists an operator $T$, making the  diagram in Fig.~\ref{commute} commutative, i.~e., such that $M^n T\gamma_n=T \gamma_0$, or more generally
\begin{equation}
M^k T\gamma_n=T \gamma_{n-k}.
\label{reversal}
\end{equation}

The operator $T$ is defined here not by the physical essence of $M$, but just by the latter requirement, and we have to \emph{find} it. If it does not exist, the map $M$ is not reversible. For reversible $M$, the inverse map $M^{-1}$ will then be, evidently, $M^{-1}=T^{-1}MT=TMT $.

\section{\label{GBM}Generalized baker map}

We will generalize the classical map~(\ref{classical}) to what may be called ``generalized'' baker map $B_w$ (GBM for short), in a way similar to the one described in~\cite{Dorfman} and denoted there as the ``slightly generalized baker's transformation''. The map is defined~\footnote{The action of $B_w$ for $1 \leq w \leq 2$ does not differ essentially from that for $w \geq 2$. In the former interval, the contraction grows towards $y=1$ as opposed to the $y=0$ for the latter. That is why we will be analyzing $B_w$ almost exclusively for $w \geq 2$.} for any $w > 1$ and is acting on points $\gamma \equiv (x, y)$ of the unit square $E=[0,1] \times [0,1]$. For the points $ 0 \leq x \leq (w-1)/w$ its action is:
\begin{equation}
B_w(x,y) = \left(\frac{w}{w-1}\: x, \frac{y}{w}\right)
\label{general_x}
\end{equation}
and for the remaining ones $(w-1)/w < x \leq 1$:
\begin{equation}
B_w(x,y) = \left(w(x-1)+1, \frac{w-1}{w}\:y + \frac{1}{w}\right).
\label{general_y}
\end{equation}

Later it will be advantageous to have the above definitions rewritten in the form
\begin{equation}
B_w(x,y) = \left\{ \begin{array}{ll}
               (L_x x, L_y y) & \mbox{ for $ 0 \leq x \leq
(w-1)/w$} \\
               (R_x x, R_y y) & \mbox{ for $ (w-1)/w < x \leq 1$,}
               \end{array}
            \right.
\label{LR}
\end{equation}
with the evident expressions for the ``left'' and ``right'' operators: $L_x x = wx/(w-1)$, $R_x x = wx-w+1$, $L_y y$ = $y/w$ and $R_y y$ = $(wy-y+1)/w$.
\begin{figure}[htb]
\begin{center}
\includegraphics[width=70mm]{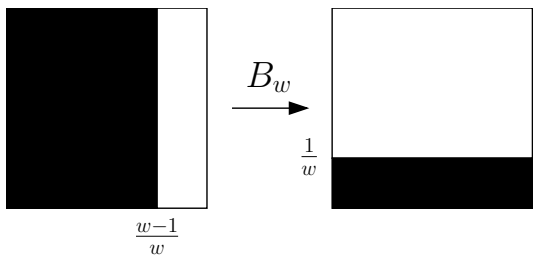}
\caption{$B_w$ transforms the filled rectangle with the area $(w-1)/w$ into the rectangle with the area $1/w$. Similarly, the complementary empty rectangle with the area $1/w$ is transformed into the one with the area $(w-1)/w$.}
\label{map}
\end{center}
\end{figure}

$B_w$ is a piecewise linear mapping (see Fig.~\ref{map}) behaving differently to the left and to the right of the vertical line $x=(w-1)/w$, which we will call the \emph{dividing line}. One can easily check that in order for the map be time-reversible, i.~e., fulfill the time reversal condition $B_w^n T\gamma_n=T \gamma_0$, one has to define the reversal operator $T$ as ``rotation'' around (or reflection with respect to) the second diagonal $y=1-x$, i.~e., as
\begin{equation}
T(x,y)=(1-y, 1-x)
\label{reverse}
\end{equation}

The map can be analyzed using modern computer algebra systems, which enable us to compute the action of $B_w$ with absolute precision. One can then observe time-reversed evolution on a computer screen, and analyze the reversibility. However, to prevent rounding errors, such computations require using rational coordinates and rational $w$, instead of finite-precision decimal values~\cite{Kumicak01}.

All simulations described in the paper were performed under the above conditions. The analysis was further simplified by restricting $w$ to integers. The latter has no effect on the results presented in the paper, so that in the following I will mostly suppose integer values of $w$.

The expansions caused by $B_w$ in the $x$ direction and contractions in the $y$ direction are characterized by local logarithmic rates $l_x$ and $l_y$
\begin{equation}
l_x = \ln \frac{\partial B_w(x,y)}{\partial x}\quad \mbox{and}\quad
l_y = \ln \frac{\partial B_w(x,y)}{\partial y}.
\end{equation}
Since the action of $B_w$ is different for points lying to the left and to the right of the dividing line, we will have two rates in the $x$ direction
\begin{equation}
l_x^L = \ln \frac{w}{w-1}\quad \mbox{and}\quad l_x^R(w) = \ln w
\label{lambda_x}
\end{equation}
and two in the $y$ direction
\begin{equation}
l_y^L = \ln \frac{1}{w}\quad \mbox{and}\quad l_y^R(w) = \ln
\frac{w-1}{w}.
\label{lambda_y}
\end{equation}

The map $B_w$ represents probably the simplest possible model exhibiting ``microscopic'' reversibility and ``macroscopic'' irreversibility, which strongly motivates its study as that of an example illustrating the foundations of irreversible thermodynamics, see, e.~g.,~\cite{Ott}. To start with, we mention that all the points of $E$ can be subdivided into fixed points, cycles (periodic orbits), and attractors, as well as points approaching those sets. Let us consider each of the sets separately.

\subsection{Fixed points}

The map $B_w$ possesses two hyperbolic fixed points, $(0,0)$ and $(1,1)$. The local stable manifold $W^s_{loc}$~\cite{Guckenheimer83} for the point $(1,1)$ contains any subset of the vertical line $x=1$ in $E$, since $B_w^n \gamma_0$ approaches $(1,1)$ for $\gamma_0=(1,y_0)$ with any $y_0$ within this subset. Similarly, the local unstable manifold for the point $(0,0)$ contains any subset of the horizontal line $y=0$ in $E$ (which we shall denote as the \emph{primary line} in view of its later role) contained in $(0, w-1/w)$, because $B_w^n \gamma_0$ departs from $(0,0)$ for $\gamma_0=(x_0,0)$ with any $x_0$ within this subset. We will discuss corresponding global manifolds $W^s$ and $W^u$ later.

\subsection{\label{Cycles}Cycles}

The generalized baker map appears to have a rich structure of periodic orbits, or cycles. Consider a point $\gamma=(x,y)$. Any combination of operators $L_x$, $R_x$ acting consecutively on $x$ will yield an expression linear in $x$, so that setting it equal to $x$ will give us an equation with a unique solution. As an example, the equation $L_x R_x L_x x$ = $x$ leads to
\[
\frac{w^3x-w(w-1)^2}{(w-1)^2} = x,
\]
with the unique solution
\begin{equation}
x_1=\frac{w(w-1)^2}{w^3-(w-1)^2}.
\label{3cyclex}
\end{equation}
We can find the solution for the second coordinate $y_1$ similarly, using the equation $L_y R_y L_y y$ = $y$,
\begin{equation}
y_1=\frac{w}{w^3-w+1}.
\label{3cycley}
\end{equation}
The point $\gamma_1 = (x_1,y_1)$ is then one of the points of the three-cycle (remaining two points can be calculated by two applications of $B_w$). Remembering that the operator $L_x$~($R_x$) acts on the point to the left (right) of the dividing line, we see that the prescribed succession of operators (we can call it the \emph{operator structure} of a cycle) will create a cycle which will visit corresponding sides of the dividing line.

There are only two structures which do not generate cycles, namely $L^n$ and $R^n$, leading to stationary points $(0,0)$ or $(1,1)$, respectively. But for any other structure there exists precisely one cycle visiting both sides of the dividing line in the given order. This shows that there is a one-to-one correspondence between the $x$ coordinate of \emph{arbitrary} point on the cycle and the sequence of positions of remaining points with respect to the dividing line. Therefore, the $x$ coordinate of a point of the cycle ``encodes'' the sequence of operators $L_x$ and $R_x$ (operator structure), and \emph{vice versa}, in a way very similar to that of the Bernoulli shift~\cite{Schuster88}. This is related to what is usually referred to as ``symbolic dynamics'' \cite{Ott}.

The set of all possible cycles has interesting structure and deserves a detailed study. Here I just mention some relevant results, starting with the remark that due to contractivity of $B_w$ in the $y$ direction, every cycle is a limit set for a subset of points in $E$.

The number of cycles grows exponentially with the cycle length. There are $2^p$ possible combinations of two operators $L_x$ and $R_x$, having the length $p$. Two of them correspond to fixed points, and in the case in which $p$ is not prime, some of the combinations (denote their number~$r$) may be further reduced to ones corresponding to shorter periods. Every $p$ of the remaining $2^p-r-2$ combinations represents cyclic permutations, so that they correspond to the same cycle. The total number of different cycles of the length $p$ is then $(2^p-r-2)/p$.

For any value of $w$ we have, consequently, an infinite number of all possible cycles. The cycles created by the same sequences of $L_x$ and $R_x$ are topologically independent of $w$, and with growing $w$ they are just scaled to smaller dimensions (the denominators are growing) and translated towards the point (1, 0). This implies that there is the same number of cycles for any $w$. In the case of $w=2$, every point with an odd denominator of the $x$ coordinate belongs to some cycle. For $w>2$, however, we do not have such a simple rule due to the above-mentioned scaling.

Consider now a period with very large $p$. Only a very small proportion out of $2^p$ sequences of corresponding operators $L_x$ and $R_x$ will be ordered in any sense --- the majority will look like random sequences. The orbits they will generate will therefore be indistinguishable from random sets of points on the trajectories.

It is evident that the set of points belonging to cycles, and of points approaching them, is of zero measure in $E$ and consequently the behavior of cycles is not what we could observe in GBM frequently.

We now come to the most important subset of $E$, which is --- as we shall see --- dense in $E$ and therefore the behavior of its points represents typical properties of GBM.

\subsection{Attractors and their properties}

Averaging the logarithmic expansion rates~(\ref{lambda_x}) and~(\ref{lambda_y}) over typical trajectories in $E$ (see~\cite{Hoover98}), one obtains what is usually called Lyapunov, or time-averaged, exponents --- the positive one
\begin{equation}
	\lambda_1 = \frac{w-1}{w}\,\ln\frac{w}{w-1}+\frac{1}{w}\,\ln w = \frac{1}{w}\,\ln \frac{w^w}{(w-1)^{w-1}}
	\label{Lyap_x}
\end{equation}
and the negative one
\begin{equation}
	\lambda_2 = \frac{w-1}{w}\,\ln\frac{1}{w}+\frac{1}{w}\,\ln\frac{w-1}{w} = \frac{1}{w}\,\ln \frac{w-1}{w^w}.
	\label{Lyap_y}
\end{equation}
From the above expressions one sees that with growing $w>2$, both exponents are monotonously decreasing, and their limit behavior for $w \to \infty$ is $\lambda_1(w) \to 0$ and $\lambda_2(w) \to -\infty$.

\begin{figure}[htb]
\begin{center}
\centerline{\includegraphics[width=70mm]{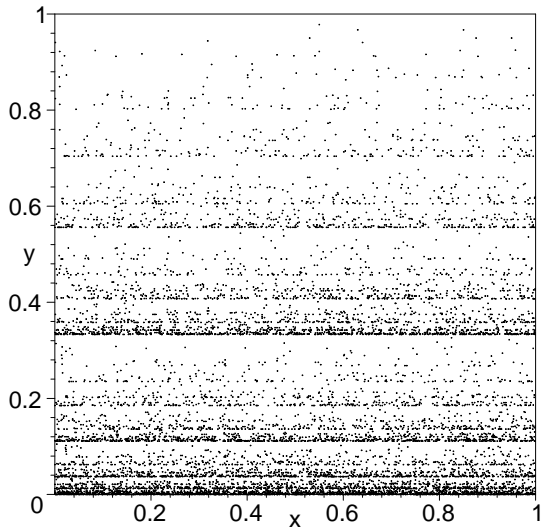}}
\caption{Iteration of the point (95/100, 95/100) by the action of $B_3$ discloses a distinct attractor with clearly visible self-similarity. 10~000 iterations are shown.}
\label{iteratedAttr}
\end{center}
\end{figure}

The existence of the positive Lyapunov exponent suggests that one should expect chaotic behavior of the iterates of $B_w^n \gamma_0$ (for almost every $\gamma_0 \in E$), and the existence of the negative one the existence of a (strange) attractor~\cite{Schuster88}. Both are observed when one iterates $B_w$ beginning with almost any starting point --- see Fig.~\ref{iteratedAttr}.

It follows from previous remarks that exceptions to this general statement include points of $W^s_{loc}$, approaching the hyperbolic fixed point $(1,1)$, and points belonging to cycles, or approaching them.

\begin{figure}[hbt]
\begin{center}
\includegraphics[width=80mm]{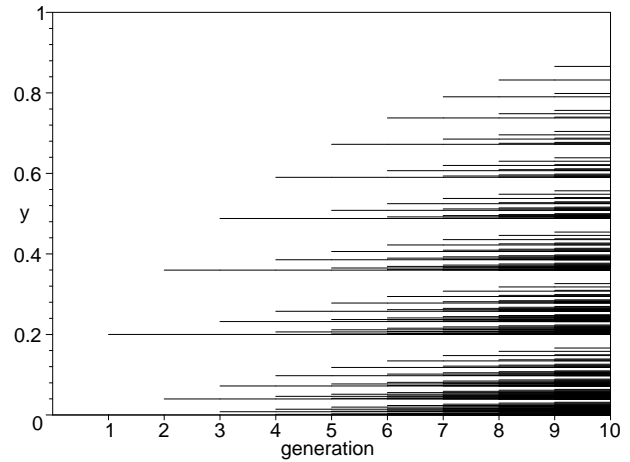}
\caption{Iterative generation of the attractor of $B_w$ ($w=5$). First, $B_w$ is applied to the line segment $y=0$ (primary line). We call the resulting two lines (including the primary line) the first generation. Applying $B_w$ repeatedly to all lines of the previous generation, we obtain $2^n$ prefractal lines after $n$ iterations. New generations are illustrated by gradually shorter lines.}
\label{generation}
\end{center}
\end{figure}

One can prove that the attractor consists of an infinite set of lines parallel to the $x$ coordinate, see Fig.~\ref{generation}, and can be generated by successive applications of $B_w$ to the primary line. The construction is based on the iterated function system, see~\cite{Kumicak01}. It is well known that in the case of $\lambda_1+\lambda_2 <0$, the attractor is a (multi)fractal object.

From the definition of an unstable manifold, it follows that the points lying exactly on the attractor (i.~e., points generated by the described construction) represent the global unstable manifold $W^u$ for the point $(0,0)$. In this sense, the latter represents the source for the attractor.

The attractor is markedly inhomogeneous (strange) for greater values of $w$. With growing $y$, its density decreases (for $w>2$, but increases for $w<2$ if we permit rational non-integer values for $w$). With decreasing $w>2$, the inhomogeneity is less and less pronounced, until at last, for $w=2$, the limit object becomes the set of equidistant horizontal lines. The latter specific case will be treated separately in Sec.~\ref{classmap}.

\begin{table}[h]
\begin{center}
\begin{tabular}{|c|c|c|c|c|c|c|c|}
\hline
$w$ & 2 & 3 & 4 & 5 & 10 & 100 & 1000\\ \hline
$D_1(w)$ & 2.000 & 1.734 & 1.506 & 1.376 & 1.156 & 1.012 & 1.001\\
\hline
\end{tabular}
\end{center}
\caption{Information dimension of the attractor of $B_w$ for some values of para\-meter $w$.}
\label{dims}
\end{table}
As with any attractor, one would like to know its dimensions $D_s(w)$, $s \geq 0$. The Hausdorff dimension~(\ref{Hausdorff}) is evidently $D_0(w)=2$ (see e.~g.~\cite{Falconer90}). The information dimension is calculated with the help of formula~(\ref{information}), taking into account that one can use, in place of probability, the density of iterates of a typical single point~\footnote{Calculation of the Kaplan-Yorke dimension $D_{KY}=1+\lambda_1/\lambda_2$ will yield the identical result.}. Repeating the calculations of Hoover and Posch~\cite{Hoover98}, but for general $w$, one arrives at the following expression for the information dimension~(\ref{information}):
\begin{equation}
D_1(w)=1-{\frac {\ln ({w}^{w}\left (w-1\right )^{1-w})}{\ln (\left (w-1
\right ){w}^{-w})}}.
\label{AttrDim}
\end{equation}

\begin{figure}[h]
\begin{center}
\includegraphics[width=75mm]{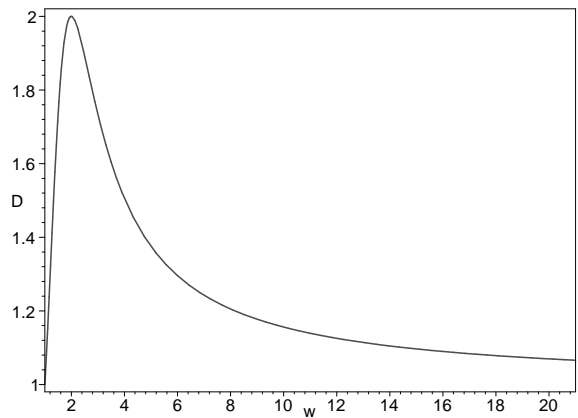}
\caption{Functional dependence of information dimension $D_1(w)$ on the value of para\-meter $w$. One sees that the dimension will approach 1 with infinite growth of $w$. The behavior for $1<w<2$ does not represent anything new since for these values the unit square $E$ is just contracted in the direction of growing $y$.}
\label{Dw}
\end{center}
\end{figure}

Table~I gives the dimension for a few values of $w$, and the graph in Fig.~\ref{Dw} enables us to see the overall dependence of $D_1(w)$ on~$w$. Evidently, one can obtain any value of $D_1(w)$ from the interval $2\geq D_1(w)>1$, by controlling~$w$. Since $D_1$ can be viewed as the measure of attractor inhomogeneity, the approach of the dimension to the value of 1, with growing $w$, suggests that the iterated points tend to accumulate (condense) on smaller subsets of $E$.

It can be proved~\cite{Barnsley88} that every attractor generated by an iterated function system is the closure of its periodic points. This suggests that the evolution generated by $B_w$ may show Poincar\'{e} recurrences.

\section{\label{Symmetry}Repellors and symmetry of evolution}

Since our main preoccupation is the study of interrelation between reversibility and irreversibility, we cannot be interested only in the ``future'' of a state defined by an initial point $\gamma_0$, i.~e., in the trajectory \emph{beginning} in this point, but we have to consider its ``past'' as well, viz.\ the trajectory \emph{ending} there. Here we will therefore try to find out what is possible to tell about the \emph{whole} trajectory going through a given point~$\gamma_0$.

Repeated application of $B_w$ to arbitrary point $\gamma_0$ generates a sequence of \emph{future} points $\{\gamma_1, \gamma_2, \dots \gamma_{n-1}, \gamma_n\}$, where $\gamma_k$ = $B^k_w \gamma_0$. Let us denote by $\gamma_{-k}$ a point from which the point $\gamma_0$  ensued after the application of the map $B_w^k$. Then the \emph{past} of the point $\gamma_0$ will be represented by the se\-quence $\{\gamma_{-n}, \gamma_{-n+1}, \dots \gamma_{-2}, \gamma_{-1}\}$. Evi\-dently, $\gamma_{-k}$ = $TB^kT\gamma_0$, and we will call the inverted sequence $\{\gamma_{-1}, \gamma_{-2}, \dots, \gamma_{-n+1}, \gamma_{-n}\}$ the \emph{backward} iteration of the state $\gamma_0$. The fact that it can be obtained also by repeated application of $B_w^{-1}$ to $\gamma_0$, is irrelevant here.

It is clear that $T\gamma_0$ is arbitrary (or random) in exactly the same sense as $\gamma_0$, so that the iterates $B_w^k T\gamma_0$ will approach the attractor as well. This has the following consequences. Since $B_w$ is contracting in the $y$ direction, the distance between any two points, having exactly the same $x$ coordinate, will quickly decrease under the action of $B_w^k$, and the properties of the future trajectory will be determined essentially by the $x$ coordinate of the initial point. The reversal $T$ interchanges the components of coordinates, so that during the backward iteration the $y$ coordinate of the original point will similarly determine the past trajectory. We can thus say that the information about the \emph{global} aspects of the future is contained in the $x$ coordinate of $\gamma_0$ and the information about the past in its $y$ coordinate. We will express this fact by writing symbolically $\gamma=(x_{fut}, y_{past})$.

This, however, means that the past of a point will approach (under backward iteration) an object symmetrical with respect to the attractor. As this object is composed of vertical lines, every point in its vicinity will move away from it under \emph{forward} iteration, since the differences in $x$ coordinates are growing with iterations. That is why it is being called the \emph{repellor}.

We have thus come to an important result: \emph{an orbit going through arbitrarily chosen point $\gamma_0$ has its past close to the repellor, and its future close to the attractor}. In other words, a typical trajectory generated by the action of $B_w$ departs (in the past) from the repellor and approaches (in the future) the attractor. Such unidirectional behavior is sometimes characterized as demonstrating the existence of the \emph{``arrow of time''} and here it is the direct consequence of the dynamics defined by $B_w$. The time reversal of such a trajectory will again be going from the repellor to the attractor, since $T$ transforms future points into the past ones (and attractors into repellors) and \emph{vice versa}. The time reversal is thus not able to change this global aspect of evolution.
\begin{figure}[htb]
\begin{center}
\includegraphics[width=70mm]{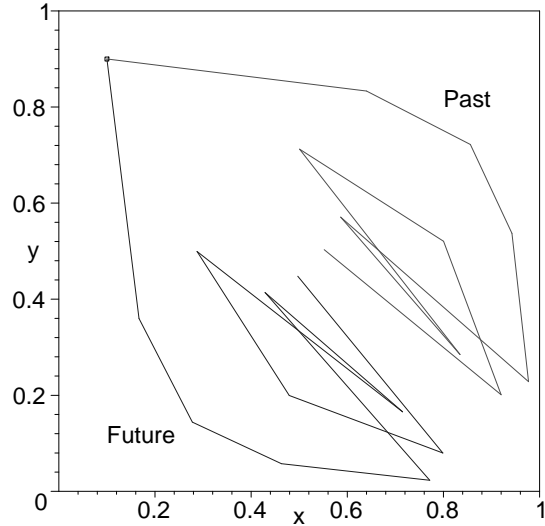}
\end{center}
\caption{Symmetry between the past and the future of the trajectory (for $w=5/2$) with the initial point $\delta_0=(1/10,9/10)$ --- denoted by a box --- lying exactly on the diagonal.}
\label{Symm}
\end{figure}

This state of affairs can best be seen when we choose a point $\delta_0$ lying exactly on the second diagonal, i.~e., a point $\delta_0=(x_0, 1-x_0)$. Such an initial point remains unchanged upon reversal and therefore its past will be the reversal of its future: past and future parts of the trajectory will be exactly symmetric with respect to the diagonal, $\delta_{-k}=T\delta_k$, see Fig.~\ref{Symm}. Or still in other words, we obtain the past part of the trajectory, in this specific case, by reversing its future.

Considering the explicit construction of the attractor, described in the previous section, we immediately see that the points lying exactly on the repellor represent the global stable manifold $W^s$ for the point $(1, 1)$. In this sense, the latter represents the sink of the repellor.

\begin{figure}[htb]
\begin{center}
\includegraphics[width=70mm]{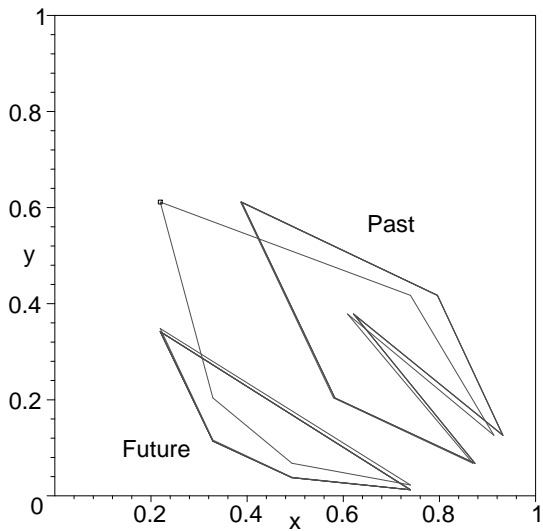}
\caption{Trajectory unwinding from a 6-cycle in the past and approaching a 4-cycle in the future (for $w=3$), ``generated'' by the initial point $(16/73,63/103)$, denoted by a box.}
\label{2cycles}
\end{center}
\end{figure}

Returning to cycles, we discover an interesting property, applying to any $w$. There are trajectories which unwind from a $p_1$ cycle in the past and approach a $p_2$ cycle in the future; see the example in Fig.~\ref{2cycles}. In my view, such behavior illustrates best the discussed difference between the past and the future, since it demonstrates that there exist two different limits in the behavior of $B_w$, separated by intermediate states. Let us mention in passing that the period of the future and the past cycle is determined --- in accordance with the general rule expressed above symbolically as $\gamma=(x_{fut},y_{past})$ --- by denominators of the coordinates $x$ and $y$, respectively.

$B_w$ can be viewed as mimicking the behavior of a thermodynamic system (a nonequilibrium thermostatted one for $w \neq 2$, and an isolated one for $w = 2$) and therefore it could be employed to explain the origin of irreversibility. In this way, one is lead to the tentative conclusion that the evolution of some physically relevant systems, sharing common properties with $B_w$, is such that their states \emph{approach an attractor} and in the past they depart from a repellor. This being the case, we arrive at an alternative explanation of the interrelation between reversibility and irreversibility, avoiding some paradoxes usually derived from the second law. For example, the question of why we do not observe the evolution going in the direction from attractor to repellor now becomes meaningless, because such a  possibility is ruled out by the system's dynamics. The problem is instead being shifted to the question of why we encounter an approach to the attractor but not the departure from the repellor --- both proceeding in one and the same direction of time. In other words, the question is, why do we observe only one part of the \emph{full} (and in principle possible) evolutionary trajectory? The next sections are devoted to tackling the problem of irreversibility in this alternative formulation.

\section{\label{absolute}Age of states}

Since the typical \emph{full} phase trajectory (considered unbounded both in the future and in the past) consists of points moving from the past repellor to the future attractor, it is quite natural to call the points between the past and the future the \emph{present} ones. This leads us to consider the possibility of introducing an ``age'' that could be applied along the trajectory.


Looking at the definition~(\ref{general_x}) and~(\ref{general_y}) of $B_w$ with rational $w$, we see that if we choose arbitrary point $\gamma_0$ with \emph{rational} coordinates, the coordinates of $\gamma_n \equiv B_w^n \gamma_0$ will be rational too, and with growing $n$ they will be represented as fractions of growing integers (except the cases of periodic orbits) --- they will become more ``complex''. The same applies to backward iteration $\gamma_{-k} \equiv TB^kT\gamma_0$. Since, by definition, the denominator of any coordinate in $E$ is not less than its numerator, we can assess the ``complexity'' or ``simplicity'' of coordinates by the value of denominators: smaller denominators will mean ``simpler'' coordinates.

Expressing an arbitrary point on the trajectory (using fractional expressions for rational coordinates) as $\gamma=(x,y) \equiv (p/q, r/s)$, we realize that in the (distant) future the mapping of the $y$ coordinates of points is dominated by $y \to y/w$, whereas that of the $x$ coordinates is dominated by $x \to x/(w-1)$. This means that the $y$ coordinates will have greater denominators than the $x$ coordinates ($s>q$), whereas in the (distant) past it will be the opposite, since the role of coordinates becomes exchanged. We could thus base the notion of ``age'' on the difference $s-q$. For future points the latter will be positive; for the past ones, negative. However, except for the points lying exactly on the diagonal (see Fig.~\ref{Symm}), the presence will not be sharp: we will not observe monotonous behavior in its neighborhood and different trajectories may have present states of different ages. This is especially true if we \emph{choose} present points with $q \gg s$, making them seem to be past in such a way.

To illustrate the situation, consider the following typical subsequence of values $s-q$ for $w=3$ and $k=-8 \ldots 8$, with initial point $\gamma_0=(1/4, 3/5)$:
\begin{center}
  \ldots, -2027, -649, -203, -61, -17, -8, -2, 1, \textbf{1}, -3, -1, 13, 103, 341, 1151, 3517, 10679, \ldots .
\end{center}
The subsequence is centered around the value 1, denoted in bold, corresponding to the initial point~$\gamma_0$.  The difference $s-q$ lacks the most important aspect of age, namely its linearity. We can obtain the latter by defining the age as
\begin{equation}
\tau=\mbox{sgn}(s-q)\log_w(|s-q|)
\label{absage}
\end{equation}
and skipping eventual (and rare) zeros for which this expression is not defined. In such a way, we get a really very precise approximation to linear age --- a result which can be understood by noticing that $B_w$ \emph{always} maps $s \to ws$, whereas $q$ either remains unchanged or $q\to(w-1)q$. Then for $\gamma_k$ we have on average $s-q \approx w^k-(w-1)^{k-1}$ for $k>0$. The $\log_w$ of the latter approaches, with growing $k$, the value of $k+const$ with very high precision. The above sample (sub)sequence is then transformed into the following one:
\begin{center}
 \dots, -6.93, -5.89, -4.84, -3.74, -2.58, -1.89, -.63, 0, \textbf{0}, -1, 0, 2.33, 4.21, 5.30, 6.41, 7.43, 8.44, \dots
\end{center}
\noindent which approaches, with growing $k$, equidistant values of ``age'', separated by exactly unit steps. One has to note that corresponding points on different trajectories can have slightly different ages.

\begin{figure}[htb]
\begin{center}
\includegraphics[width=55mm]{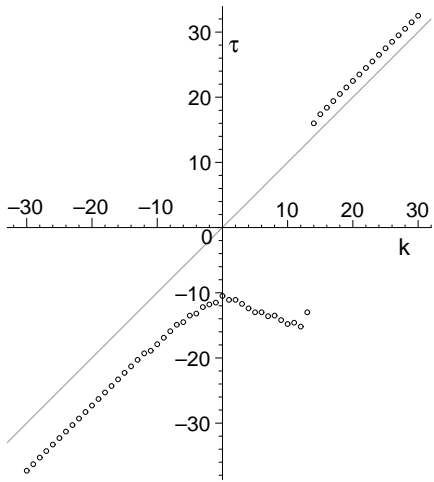}
\caption{Dependence of ``absolute age'' $\tau$ on iteration index $k$. Although one of the worst cases, with initial point (50~001/100~000, 11/30), has been chosen here intentionally, the age $\tau$ differs from $k$ by less than 10. This situation is far from typical, because in a random choice of a point it is not probable to obtain such a great difference in the precision of its two coordinates. The typical behavior is characterized, contrary to the depicted one, by two parallels close to $\tau(k)=k$, exhibiting a sudden jump around $k=0$.}
\label{age}
\end{center}
\end{figure}

These reasonings demonstrate that in a microscopically described system of GBM, the age of any point can be found in an  unambiguous way and with reasonable precision; see Fig~\ref{age}. It will increase in equal steps from past to future --- except the small neighborhood of the ``present'' state. This age is \emph{absolute} in the sense that it does depend only on the distance from the present, i.~e., on the number of iteration steps. This is due to the fact that coordinates of points contain ``traces'' of their evolution, which enable us, at least in principle, to recover the iteration index $k$ from the coordinates of arbitrary point. It is appropriate to stress here that the case of the classical baker map ($w = 2$) is no exception to this general rule.

The described approach to age might seem extremely artificial, but it can be viewed, nevertheless, as representing what may be called ``microscopic'' age. Of course, one should not overemphasize this model description, but neither should it be underestimated. In my view, it demonstrates --- together with the above-mentioned behavior of cycles, illustrated in Fig.~\ref{2cycles} --- that there is, at least \emph{qualitatively}, a difference between past and future already at the microscopic level. This then implies the existence of the property of the system, which I would call \emph{microscopic irreversibility}.

If we now consider the evolution of a set of $N$ points under the action of $B_w$, the above reasonings will apply to any point of the set. We could therefore --- at least in principle --- define in any evolutionary trajectory of the set the ``age'' as the average of ages of all individual points. The ``present'' state would then be the one with the (averaged) age closest to zero.

However, looking at the set without the knowledge of the coordinates of individual points, it is possible to know its ``age'' also by observing its shape. Specifically, in the case of a subset of $E$ having a simple boundary (e.~g., the case of points lying within a small square), the $B_w$ --- causing the growth of fractional expression of coordinates of all the points --- will cause also the ``distortion'' of the boundary. This then means that such distortion is commensurate with the age of the set. To make this idea more definite, we turn to the distribution functions on $E$, giving the density (of probability) of points in $E$.
\begin{figure}[h]
\begin{center}
\includegraphics[width=70mm]{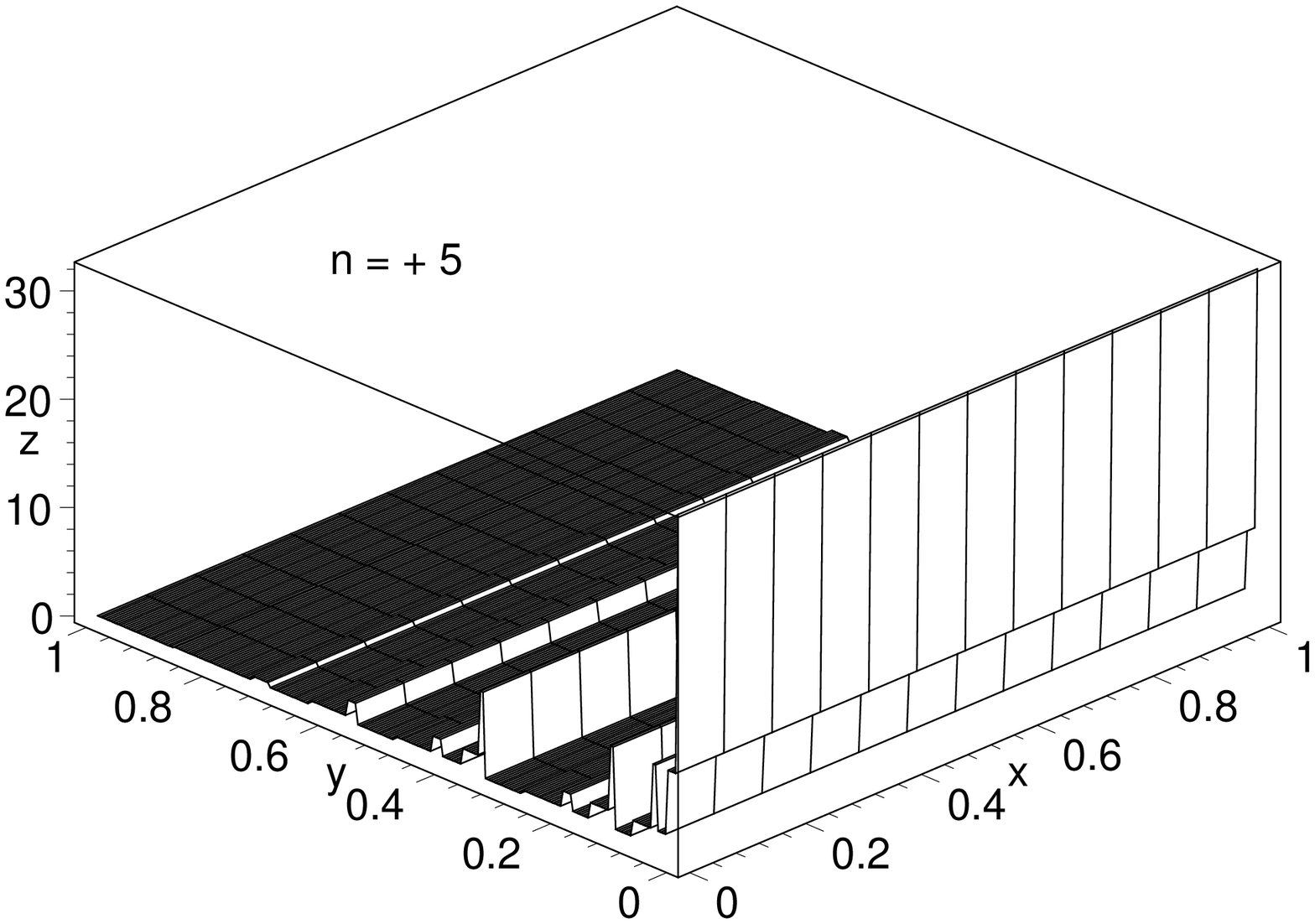}
\includegraphics[width=70mm]{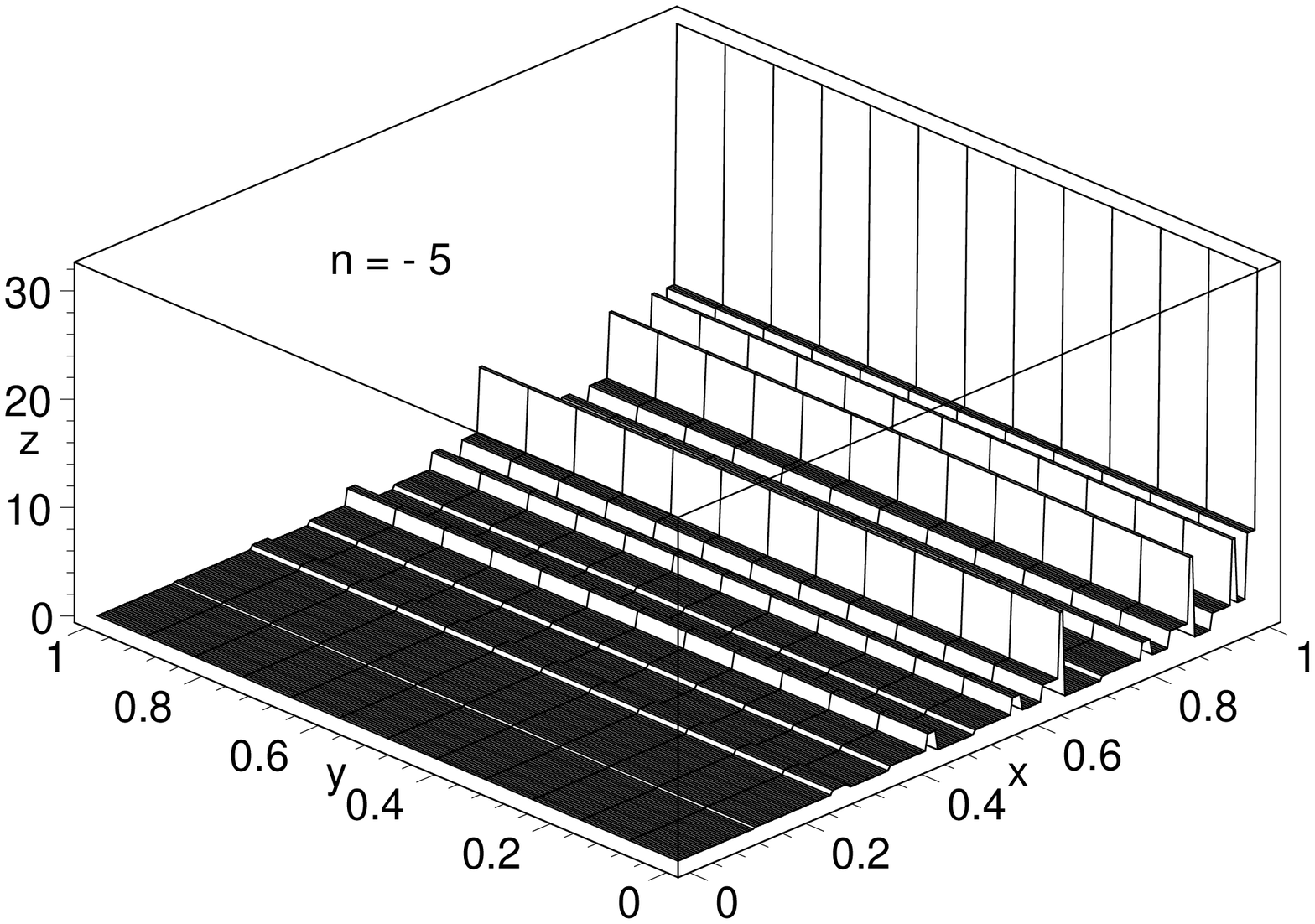}
\caption{Two states in the evolution of a distribution function (under the mapping $B_3$), the present state of which is defined as constant on the unit square.}
\label{3D}
\end{center}
\end{figure}

\begin{figure}[htb]
\begin{center}
\includegraphics[width=75mm]{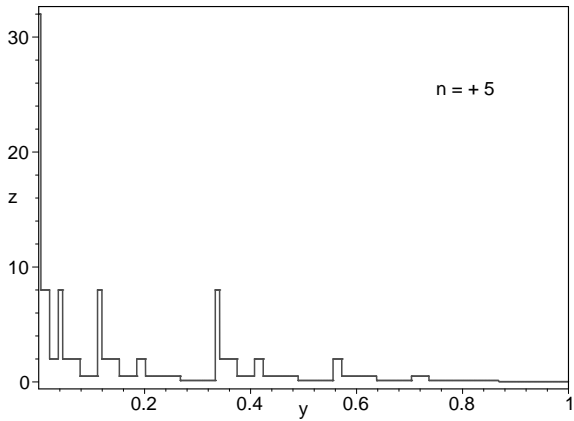}
\caption{Section across the distribution function $\phi_5 \equiv B_3^5\,\phi_0$, with the initial function $\phi_0(x,y) \equiv 1$, by a plane perpendicular to the $x$ axis. The maximum value is $\phi_5(0,y)=32$.}
\label{Cross}
\end{center}
\end{figure}

The map defined by $B_w$ on $E$ will translate in a straightforward way into the evolution of distribution functions having $E$ as their support. The distribution function, constant on a compact subset of $E$, will in the forward evolution approach a function independent of $x$, and in the backward one, independent of $y$ (see Fig.~\ref{3D}). But even general distribution functions will very quickly approach functions almost independent of $x$, or $y$. That is why one can characterize, with sufficient precision, the \emph{future} limit behavior of any distribution function with the help of the function $f(x_0, y)$, representing a section across $f(x, y)$ by a plane perpendicular to the $x$ axis, going through arbitrary point $x_0$, see Fig.~\ref{Cross}. For such a function of one variable, a variation on the unit interval is defined as the supremum
\begin{equation}
V_y(f)= \sup \sum_{j=1}^m |f(x_0, y_j)-f(x_0, y_{j-1})|,
\end{equation}
taken over all possible dissections of the interval by $m>1$ points. We can regard this expression as an acceptable approximation of the variation of the distribution function along the $y$ axis.

In quite a similar way, one can characterize the \emph{past} limit behavior of any distribution function with the help of the variation $V_x(f)$ of the function $f(x, y_0)$, representing a section going through arbitrary point $y_0$. The difference $V_y(f)-V_x(f)$ of variations along $y$ and $x$ will then grow steadily for $n$ from $-\infty$ to $+\infty$, and we can use it to estimate the ``age'' of the state, similarly as we used the difference of denominators in the case of a single point.

This then gives us the general picture of the evolution of a distribution function under the action of the map $B_w$. The absolute value of its variation will grow indefinitely for $|n| \to \infty$, attaining somewhere in between its minimum. Choosing one of the states having low variation for the ``present'' state and denoting the corresponding iteration index by $k=0$, we will again introduce what may be called the ``age'' $k$ of the system in question. We could also go a bit further and make this age linear by taking the logarithm of the difference of variations, similarly as in the case of a single point. Such an age would then be mathematically well defined everywhere, with the possible exception of the neighborhood of the ``initial'' distribution function.

Let us note in passing that the distribution function that evolved from standard ``testing'' function $\phi_0(x,y) \equiv 1$ will have the future maximum values (for $w > 2$) close to the line $y=0$ and the minimum ones close to the line $y=1$. After $n$ iterations, the values will be $(w-1)^n$ and $(w-1)^{-n}$, respectively. This can be readily seen by taking into account that the area $(w-1)/w$ to the left of the dividing line is contracted by the action of $B_w$ to the area $1/w$, i.~e., by the factor of $w-1$. In a similar way, one will understand the behavior of the minimum.

The ratio of the maximum to the minimum value is then $(w-1)^{2n}$, which becomes --- for $w>2$ --- a huge number after just a few iterations. This illustrates very well the growing inhomogeneity of the distribution functions under the action of GBM.

\section{\label{Irrev}The problem of irreversibility}

Having introduced the notion of \emph{absolute age} in a way formally different from the one used by the Brussels school~\cite{MPC79, AntoniouMisra91}, but conceptually --- as it seems --- rather close to it, we are ready to explain the apparent contradiction between microscopic reversibility and macroscopic irreversibility of generalized baker maps, or in other words, to explain the essence of the second law of thermodynamics as applied to this simple model.

The so called ``problem of irreversibility'' is usually expressed in two distinct formulations: one concerns the evolution of a state prepared by an experimenter, the other applies to the observation of a system without any question about its origin, i.~e., of a system with an unknown past. We are thus facing two questions (both equivalent to the Loschmidt objection) which are being posed since the first explanation of the second law by Boltzmann in 1872: if both directions of evolution are microscopically equally possible, then (a)~why we are not able to \emph{prepare} states which would evolve in a direction ``prohibited'' by this law, and (b)~why do we never \emph{observe} such evolution in systems with an unknown past? In the language of the absolute age, these questions reduce to the following ones: why are we not able to prepare, nor to observe, states which belong to the \emph{distant past} on the evolutionary trajectory of the system under consideration?

We will treat the two questions separately, because the explanations will turn out to be different for the two cases. We will also treat separately two different approaches, corresponding to microscopic and macroscopic descriptions.


\section{Microscopic irreversibility}

\subsection{Irreversibility in evolution of a single point}

We have shown that if we choose a point (with rational coordinates) in $E$ in any way, the coordinates of its future and past iterations will be almost always (i.~e., with the exception of zero measure of points) more complex than the present ones. The claim that we can prepare only absolutely ``present'' points will then become actually a self-evident \emph{tautology}, stating that we can prepare ``simple'' coordinates more easily than the ``complex'' ones. This tautology, however, expresses exactly the essence of the problem. If we would ask why it is not possible to prepare an ``old'' point, the coordinates of which would become simpler in the future, we could first of all point out that if that were even possible in principle, such simplification would last only for a finite number of iterations --- until the coordinates would become relatively simple (the state would become ``present''). However, to prepare \emph{ab initio} a point close to the repellor, i.~e., a point with specific and thus very complicated coordinates, means to prepare an exceptional point, and this is essentially more difficult than to generate a point at random. I propose to call this state of affairs, in the given context, \emph{microscopic irreversibility}.

It is evidently possible to prepare such a point, even numerically, only by choosing a point $\gamma_0$, letting it evolve to the point $B_w^k \gamma_0$, and at last inverting the latter. The point $T B_w^k \gamma_0$, having the age of $-k$ iterations, would then approach, under the action of $B_w$, the point $T \gamma_0$ and the coordinates of its $k$ iterations would be simplified. But it is not possible to find a general construction nor formula defining a set of such ``old'' points.

From this point of view, the microscopic irreversibility is related to substantially different requirements for preparation of the present and past states, respectively. Evolution leaves ``traces'' on the coordinates of the evolved point, and inverse evolution would require us, even in the case of a single point, to reconstruct precisely its whole history. But nothing similar is being required to prepare the present state. This crucial asymmetry lies, in my opinion, at the root of the explanation of irreversibility~\footnote{It is equally difficult to prepare points belonging to a \emph{given} distant future, but the law does not apply to this.}. One should not, however, forget that this situation is the result of chaotic properties of the evolution induced by the map $B_w$. If the map did not lead to chaotic dynamics, there would be no tautology; nothing would be self-evident in this sense.

We assumed that phase points have rational coordinates. The considerations of the last two paragraphs hold, however, equally for irrational coordinates. In particular, the fact remains that inversion of evolution would require us to recover exactly its whole history. In this sense, the argument applies to any coordinates.

Here we come to a variance with traditional view of irreversibility as formulated, e.~g., by Bricmont~\cite{Bricmont96}: ``All the familiar examples of irreversible behavior involve systems with a large number of particles (degrees of freedom). If one were to make a movie of the motion of one molecule, the backward movie would look completely natural.'' Certainly, playing back the movie of a \emph{single} point moving under the action of GBM, we would not be able to distinguish the backward movie from the forward one. But why should the physical properties depend on our common way of observation or on our observational possibilities? And what \emph{finite} number of points is sufficient to distinguish between forward and backward motion? Superimposing all the phase points of the trajectory of a single particle onto one frame, as in Fig.~\ref{iteratedAttr}, we would be able to make the distinction between the past and the future, between impossible and natural.

The simple model system of $N$ noninteracting particles in a rectangular box~\cite{Kumicak77} shows that large number of particles alone is not sufficient for irreversibility. The arguments of this section support this line of thought in that the dynamics of some systems may bring about the
unidirectionality of microscopic evolution, manifesting itself in the appearance of attractors (and repellors). Observing then systems of $N$ particles in the usual way, the inherent microscopic irreversibility becomes more and more pronounced with growing $N$, but it is present in the system all the time.

Moreover, the approach connecting irreversibility with a large number of particles is not straightforward either: it requires, e.~g., thermodynamic limit $N \to \infty$ to obtain irreversibility. For further details, see, e.~g.,~\cite{CohenRondoni98}, giving a nice treatment of the difference between chaotic and irreversible behavior of Hamiltonian systems.

It might be appropriate to mention at this point the Poincar\'{e} recurrence theorem, stating that any isolated, locally measure-preserving system will eventually return close to any of its previous states. Approach to an attractor, an object dense in $E$, has such a recurrent property. But the property of \emph{being close} in phase space does not mean being of a \emph{similar age}: in an arbitrary close neighborhood of any point, there can be points of any age. The mixing property of $B_w$ implies that the distance in the sense of metrics does not imply the distance in time, nor \emph{vice versa}. In this sense the recurrence does not contradict irreversibility.

\subsection{Irreversibility in evolution of many points}

The analysis of behavior of a single point may be of interest on its own, but the essential properties of the model are seen much better when observing the action of GBM on a set of points.

\begin{figure}[htb]
\begin{center}
\includegraphics[width=55mm]{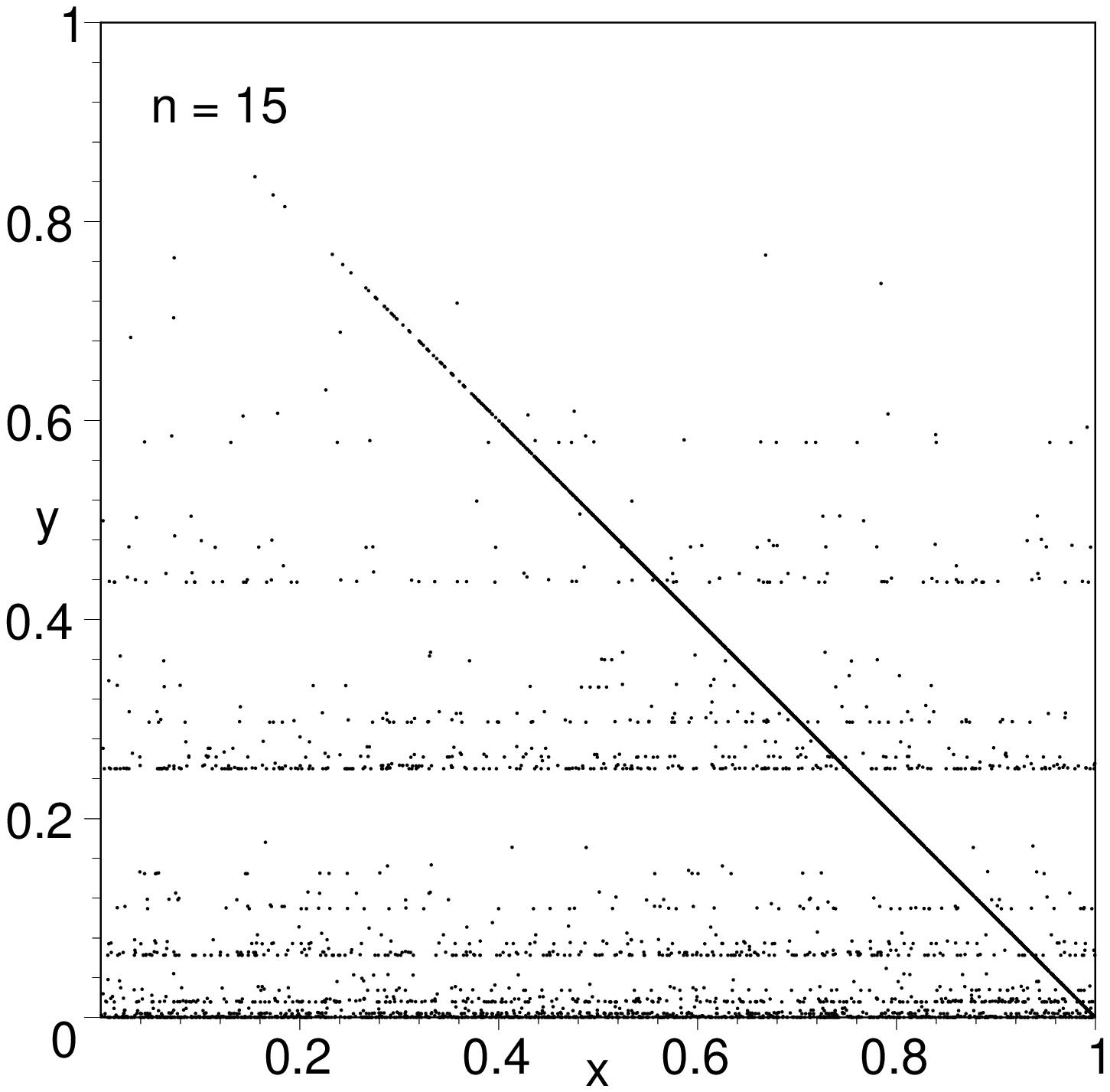}
\includegraphics[width=55mm]{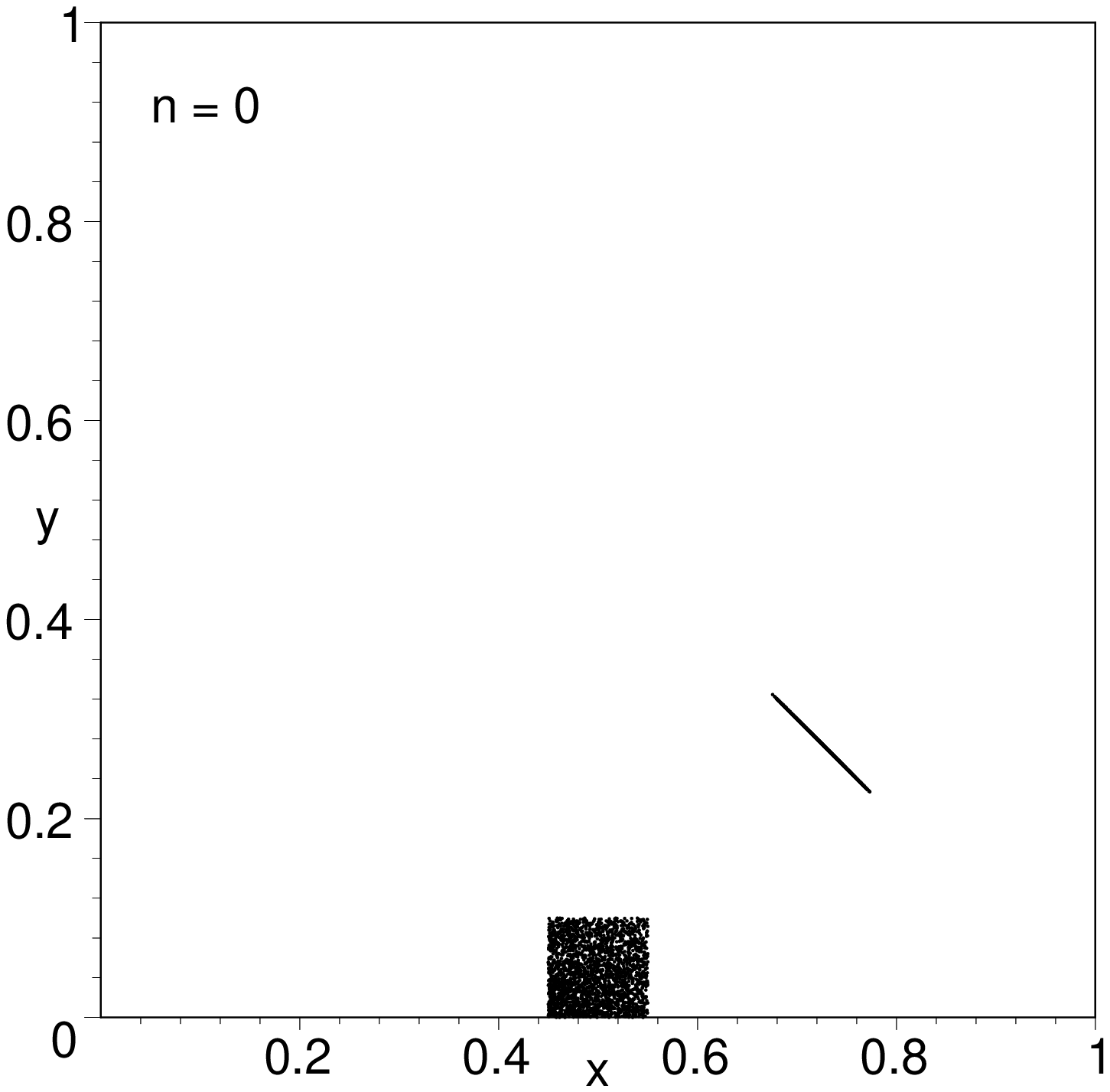}
\includegraphics[width=55mm]{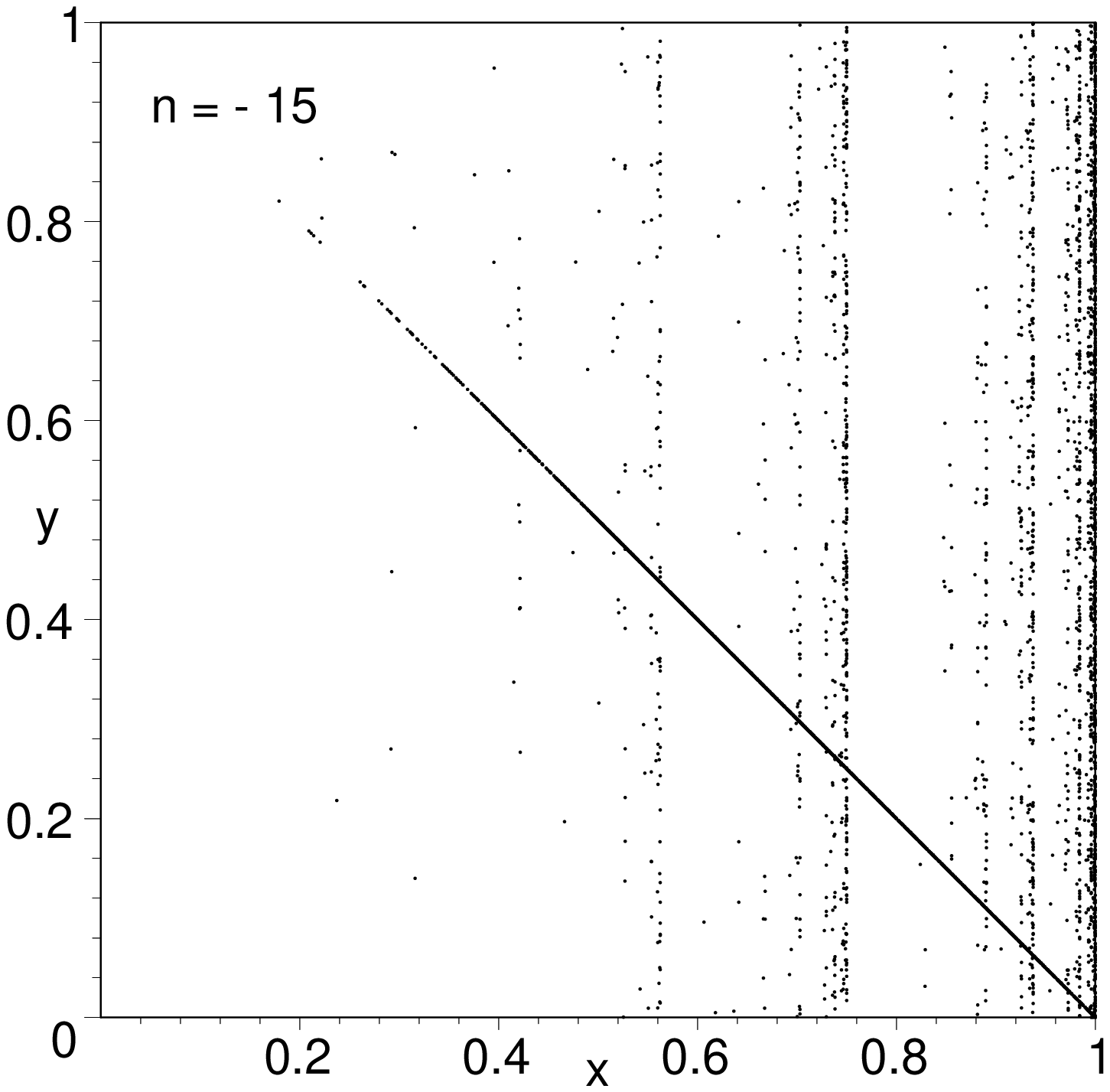}
\caption{Three states in evolution (with $w = 4$) of an initial set of 2500 randomly chosen phase points, concentrated in the square subset of $E$. We are readily able to distinguish between the past, the present, and the future. The projections of points on the diagonal $y = 1 - x$ represent an analogy to the positions of physical particles.}
\label{Square}
\end{center}
\end{figure}

The initial state can be, in the case of GBM, prepared essentially by defining a subset of $E$ and filling it uniformly with random points (or according to an appropriate distribution function). The evolution of such a set can be observed in the author's animation~\footnote{The animation can be found at the web addresses \textsf{http://www.sjf.tuke.sk/ket/kumicak/BakAnima.html} and \textsf{http://www.geocities.com/kumicak/BakAnima.html}}, where the subset can be chosen as a rectangle. Watching the evolution, one can see that points contained originally in the bounded subset will in the future get into a growing number of horizontal stripes, and in the past into vertical ones. Going from the past to the future, they will pass through the ``present'' subset. This situation is depicted also in Fig.~\ref{Square}.

Consider a set $\Omega_{-k}$ of $N$ points from which a set $\Omega_0$, located within a small square, will evolve after $k$ iterations. The greater the $k$, the closer $\Omega_{-k}$ is, in the usual sense of metrics, to a set $\Omega_{r}$ of points randomly distributed over (the repellor in) $E$. However, irrespective of how close the latter two sets become, their future behavior will be essentially different. After $k$ iterations, the first will become $\Omega_0$, the second will look like the attractor.

The first question pertaining to the second law can be formulated in this case as asking why it is not possible to \emph{prepare} a state corresponding to a distant past. To answer it, it is sufficient to note that the evolution of a set of points, although being reversible, does not offer an equal possibility for preparation of all initial states. Present states --- e.~g., points filling the small square in $E$ --- can be prepared without the slightest problems, but to prepare a state $n$ steps before would require localization of points in $2^n$ vertical stripes of $E$, and in every stripe with different, and well specified, density (cf. Fig.~\ref{Cross}). How impossible this is can be best understood when we realize that this task is equivalent to the one of \emph{preventing} the occurrence of points in the complementary subset.

Then the answer to the above question is that preparation of past states would require us to create them according to a distribution function with wild variation, which is \emph{practically} impossible. Moreover, this function is extremely close to the other one, obtained by averaging of the former, and describing a readily preparable present state. So the preparation would have to be, in addition, extremely precise. Here again we see the crucial role of a sensitive dependence on initial conditions, since two distribution functions, differencing ``macroscopically'' only slightly, have an essentially different future.

The second question, why we do not \emph{observe} states typical for the distant past, can also be answered easily: we do not observe them because any such state quickly becomes ``present'' or future, and observation means that we are watching a system with an unknown past without influencing it.

\section{Macroscopic irreversibility}

Let us find out now what we can learn about GBM if we want to consider it as a model of a physical system, exhibiting the difference between microscopic and macroscopic behavior. The difference derives typically from the distinction between the full description in phase space and various reduced descriptions. In the case of GBM, the unit square $E$ can be viewed as the phase space of one particle (denoted usually as the $\mu$-space), and any set of points in it as the complete description of a state of a 1D system of noninteracting particles. If we want to use the model to illustrate the micro/macro dichotomy, we have to find a counterpart to the less complete description.

We depart from the inversion operator $T$, the effect of which is analogous to the inversion of momenta in classical mechanics. If the GBM has to resemble, at least remotely, the dynamics of 1D gas, then the particle position should be represented by a variable which is an invariant of $T$. The projection of points in $E$ onto the ``second'' diagonal, $y = 1 - x$, has this property and we therefore assume that it represents positions. The projection of the attractor looks like a stationary (non)equilibrium distribution of points, which can be called the \emph{``limit state.''} Due to the symmetry with respect to the diagonal, the projection of the repellor will be \emph{identical} to that of the attractor. The orthogonal projection of a point in $E$ onto the ``first'' diagonal, $y = x$, could be viewed, with certain reservation, as the ``momentum'' of a particle.
\begin{figure}[h]
\begin{center}
\includegraphics[width=65mm]{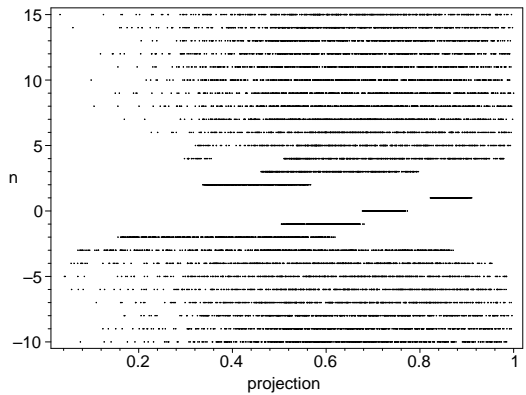}
\caption{The projection of evolution depicted in Fig.~\ref{Square} onto the second diagonal (scaled to unity length). Only 500 points were used to make the structure visible. The left-hand side of the line corresponds to the point $(0, 1)$ in $E$. We are evidently not able to distinguish between the past and the future of the projected set, but we readily see the difference between both on the one hand, and the ``present'' on the other.}
\label{Diag}
\end{center}
\end{figure}

The ``macroscopic'' behavior of the system of particles is then described by what can be seen on the second diagonal (see Fig.~\ref{Diag}). There we no longer observe the approach to the attractor, nor departure from the repellor. In the direction from present to future, we see only the change of inhomogeneous distribution of positions into the time-independent limit state. In the direction from past to present, we see the opposite change of almost homogeneous (non)equilibrium state into the inhomogeneous present one. Moreover, we are not able to distinguish between the past and the future due to the attractor/repellor symmetry. The transition from past to present contradicts the second law for a gas on a line; the transition from present to future is in accord with it.


We can now answer the two questions posed traditionally with respect to the second law. We already know that we cannot \emph{prepare} microscopically the ``past'' state because this is practically impossible. But the macroscopic preparation offers still fewer possibilities, since in this case we can ``assign'' only positions to particles, but not ``momenta''. If we were even somehow able to create some specific arrangement of points on the diagonal, we would know nothing about their ``momenta'', so the latter would be necessarily random, and consequently the phase points would not be on the repellor. This explains why it is not possible to prepare macroscopically a state which would evolve contrary to the second law, and it seems to be the full explanation.

We cannot \emph{observe} evolution going from the repellor to the present state for different reasons. First, any observation of a closed (isolated or thermostatted) system has finite duration. If we regard the trajectory generated by $B_w^k$ as a sequence of states with \emph{unbounded} past and future, then the probability of observing any \emph{specific} finite sequence of states is zero. The observation of the present state is by definition specific and finite, so the probability to find it at random (on a randomly chosen trajectory) is vanishingly small. But the probability of picking at random \emph{arbitrary} interval with exclusively past states (on the repellor) is the same as that of seeing one with exclusively future states (on the attractor) --- i.~e., 1/2. However, we cannot distinguish macroscopically between the past and the future states. This explains why we are not able to know that we have seen a system in the past even if it would be there. This holds even under the assumption that systems can exist under unchanged outer conditions indefinitely.

But the trajectories of closed systems do not have unbounded past. In the finitely remote past a state ``came into existence'' as a result of interaction of the system with the surroundings, and afterwards its boundary conditions remained unchanged. That ``first'' state was actually the ``present'' one, and from the moment of its appearance the system tends to the attractor. Therefore, the unbounded past of the presently observed state is actually an illusion: from the existence of the present state we can \emph{assume} the existence of the past states, we can even in principle recapture them, but the mere existence of the present state does not guarantee the real existence of past states on the repellor. Thus we can find the system always only in the future, approaching the attractor. This explains the second law in the form stating the properties of observed states, because it describes the behavior of a system with an unknown past.

This leads us to formulate the macroscopic irreversibility of the GBM as meaning that the evolution of this system approaches an attractor, and if it would be possible to recover its evolution from the past, the system would depart from the repellor. Such formulation is not subject to the Loschmidt paradox; it is free of any seeming contradictions. In my view, these are sufficient reasons to accept it. However, to do this, we have first to take into account that the usual statement of irreversibility applies to conservative (isolated) systems which should be simulated by $B_2$ --- the classical baker map.

\section{\label{classmap}The case of the classical baker map}

In the previous reasonings, we were dealing mostly with the full interval of parameter values, $w > 1$. However, the specific case of $w = 2$, corresponding to the classical baker map, does not share all general properties stated above, and the question therefore arises as to which of the properties do not carry over to this specific case and what consequences this may have.

The attractor of the GBM reduces, for $w = 2$, to the set of horizontal lines with $y_{ik} = c_i/2^k$, where $k$ denotes the generation number and $c_i < 2^k$ is an integer. Similarly, the repellor becomes the set of vertical lines with $x$ coordinates of the same type, $x_{ik} = c_i/2^k$. Speaking mathematically, these sets represent an attractor/repellor pair as well (see e.~g.,~\cite{Barnsley88}, p. 82), therefore we will continue to denote them so.

From the physical point of view, however, those objects are usually considered irrelevant, since they are not \emph{``strange''}. All dimensions~(\ref{dimension}) of the attractor and repellor namely reduce to~2, so that the latter are \emph{homogeneous} and consequently do not represent readily ``visible'' objects. Moreover, the sum of Lyapunov exponents
\begin{equation}
\lambda_1 + \lambda_2 = \frac{2-w}{w} \ln(w-1),
\end{equation}
which should be negative for strange attractors, becomes now zero, which means that the measure of subsets of $E$ is not being contracted under $B_2$. Such a homogeneous attractor, nevertheless, ``attracts'' the points iterated by $B_2$ in the important sense that the iterates visit the neighborhood of any point of $E$ with \emph{equal} probability.

Returning now for a while to cycles, we have observed that the number of periodic orbits containing points with denominator not exceeding a given number $d$ grows with falling $w \geq 2$, until in the case of $w = 2$ \emph{all} points with a rational $x$ coordinate, differing from those of the attractor/repellor ones, lie on trajectories approaching cycles. This can be understood considering that the $B_x$ reduces in this specific case to $x\to 2 x \mbox{ mod }1$ and the condition of a $p$ cycle, $B^p_x x = x$, then has a unique solution for any rational $x$. The attractor is therefore approached only by points with irrational $x$ coordinates, and as the former is uniformly dense in $E$, this behavior can be understood as chaotic approach to homogeneous coverage of $E$. But the approach to very long cycles is indistinguishable from the approach to a homogeneous attractor, so that also the majority of trajectories with rational points will seem to approach the latter. The typical observable behavior of all points --- rational and irrational --- will then look like chaotization, or approach to equilibrium.

Accepting the terminology according to which the attractor (repellor) exists even in the classical baker map, being only ``invisible'' there, we can formulate the obtained results in a unified manner, treating classical baker map, caricaturing an isolated system approaching equilibrium, in the same way as thermostatted systems approaching nonequilibrium steady states.

The notion of absolute age has the same meaning for $w = 2$ as for $w > 2$. The present states are defined by simple coordinates, and the present distribution functions have small variation. One can tell the difference between past and future in the evolution of a point, and choosing an inhomogeneous initial (present) distribution function, one can distinguish microscopically between past and future not only by absolute age but also by predominantly vertical or horizontal stripes of maxima, respectively --- at least in the initial stages of evolution, not too far from the present states.

Absolute age is essential, in my view, for the explanation of irreversibility. What was told about it above applies equally to the specific case of $w = 2$ and supports the view that what makes the difference between reversibility and irreversibility is actually observability: \emph{macroscopic} irreversibility is observable, whereas \emph{microscopic} is not --- it is hidden for macroscopic methods of observation. In the case of $w = 2$, it is hidden also for microscopic methods. Only the generalization to $w \neq 2$ is able to disclose its microscopic existence. And that was the reason for the study of the generalization of locally measure preserving maps.

In this context, we should not leave unnoticed the Zermelo paradox, relying on the Poincar\'{e} recurrence theorem. According to it, even if a nonequilibrium state of an \emph{isolated} system evolves to equilibrium, it must eventually return close to the initial state, thus violating the second law. This means, as applied to $B_2$, that a system of points, starting e.~g., in a small subset of $E$, will eventually return close to it. Apart from the standard reply that the recurrence time is enormously long, the model of $B_2$ enables us to give further refutation of the objection. Since all rational points approach cycles, such recurrent behavior, albeit not excluded, would be possible only for extremely exceptional sets. Choosing namely the same denominator of $x$ coordinates of all points, one obtains --- as a rule --- the sets ``condensing'' on a limited number of points in $E$. However, it is not too difficult to find also points, which will approach periodically moving sets. This confirms the existence of recurrences, but also demonstrates how rare and physically irrelevant they are~\footnote{To give a simple example, the denominator 397 will yield, for all points in the subset $x<1/10$, the set of 27 points returning to the subset every 44 iterations.}. In this sense, the recurrence does not contradict the formulation of the second law about the arrow of time pointing \emph{all the time} from the past to the future.

\section{From the simple model to real-world physics}

I believe that the model of generalized baker map is interesting even in itself. However, it was studied with the hope that its analysis will shed light on the properties of more realistic physical model systems. By the latter I mean mathematical models, the essential aspects of which mimic the properties of corresponding real physical objects. Let us start, therefore, the concluding discussion by reconsideration of the basic assumptions used in the study of the generalized baker map.

All my numerical simulations were restricted to rational coordinates. The reason was to circumvent artificial irreversibility, which would otherwise appear as a numerical artifact due to rounding errors. Rational numbers are a dense subset in the field of real numbers, and almost everything mentioned above for rational coordinates of points will apply to real ones as well. There is actually only one major problem which the exclusive use of rational numbers could cause in our study: the absence of periodic orbits for trajectories containing points with irrational coordinates. This to me does not seem to be an essential problem, if we are aware of it.

Many ``realistic'' (in the above sense) $N$-particle model systems (e.~g., dilute gas models with Lennard-Jones~\cite{Stoddard73} or hard-sphere~\cite{Dellago96} potential) share a sensitive dependence on initial conditions with GBM. $B_2$ is known to be strongly mixing~\cite{Falconer90} and the same is confirmed for many classical systems~\cite{Sinai00}. The mixing property is a consequence of Lyapunov instability, discovered in many interesting models~\cite{Dellago96}. The evolution of systems with this property leads to local contractivity of phase volume (measure) in the forward direction of time (implying the opposite when tracking the evolution backwards), or to uniform coverage of phase space by the trajectory. Approach to an attractor in GBM is then analogous, in these more realistic systems, to approach to nonequilibrium steady state, or to equilibrium.

The notion of absolute age enables us to view the past and the future as related to (strange or homogeneous) repellors and attractors --- the structures in phase space, defining the arrow of time. The age, as introduced here, relies on the property of $B_w$, that the two coordinates of phase points behave differently under $B_w$. It is evident that this results from different contractivity of the map in different directions. We can therefore expect some signs of the absolute age in all systems with strange or homogeneous attractors.

We pointed to the different role of positions and momenta in the transition to macroscopic description.
The momenta are responsible for the microscopic difference between past and future steady states, but they do not help us in their macroscopic identification. This state of affairs lies at the basis of macroscopic indiscernibility between past and future and thus supports our view that macroscopic description is intimately connected to the loss of information about the system under consideration~\cite{KumicakBrandas87a}. This may best be seen in the fact that we are readily able to impose macroscopic constraints on positions, but we are not able to do the same (or at least not to a comparable extent) with momenta. This dramatically reduces our ability to influence the states of physical systems, thus leading to apparent irreversibility.

In this sense, the GBM manifests features which coincide with the general properties of e.~g., rarefied real gases, as regards the global behavior in the distant past and future --- as far as the  interrelation between reversibility and irreversibility is concerned. The ``mechanisms'' by which irreversibility appears in reversible systems are common to GBM and strongly mixing models of $N$-particle systems. We may thus say that GBM represents an appropriate --- albeit extremely simple --- model, demonstrating essential aspects of interrelations between reversibility and irreversibility. The properties which became apparent in the study of GBM are therefore applicable to more realistic models, as follows.


In real experiments, we prepare a state by some manipulations with particles. The mathematical counterpart of such preparation is the localization of phase point $\omega_0$ of a realistic system in its phase space. The evolution of the state then results from physical laws and is \emph{represented} as the action of mathematical operator $S_t$. Whatever the physical preparation of a state may be, the localization will always result in a set of numbers having absolute precision. Of course, we do not know any state with infinite precision, but from the viewpoint of classical mechanics,  any state \emph{must} be absolutely precise \emph{per se}. Application to $\omega_0$ of $S_t$, simulating relevant physical law, yields absolutely precise numerical results for $\omega_t$, for any value of $t>0$, provided the system is deterministic.

As far as GBM is concerned, one can observe (on a monitor) even evolution going from repellor --- i.~e., from the distant past --- to the present. Such evolution will take only finite number of iteration steps: after reaching $\omega_0$, the point will move (forever) to the attractor. However, we cannot reconstruct past evolution in real physical experiments, since we cannot prepare a state corresponding to the precise localization of the point $T \omega_t$. A notable exception is provided by experimental manipulations such as those used in spin-echo experiments~\cite{Hahn53}, mimicked in numerical experiments by mathematical reversal of future states. However, due to the sensitive dependence of $\omega_t$ on $\omega_0$, the slightest deviation from the exact value of $T\omega_t$ would cause a completely different trajectory.

The same applies to the preparation of states according to a distribution function. In both cases, we prepare states and functions which are ``present'' and we are not able to prepare ``past'' states nor functions, which would eventually evolve into the present ones. Our possibilities to prepare states of physical systems are rather crudely restricted to states which are in the absolute sense ``present''.

In my view, this explains the seeming contradiction between microscopic reversibility and macroscopic irreversibility as rooted already at the microscopic level. Or still in other words, the problem of irreversibility has \emph{primarily} nothing to do with the micro/macro dichotomy. It is connected to the past/future dichotomy instead, manifesting itself in the asymmetry of preparation of states close to the repellor/attractor. The macroscopic level just adds a further restriction, consisting in our impossibility to observe or practically influence momenta of particles.

The view presented in the paper may seem to contradict the orthodox approach, stressing the role of a large number of particles in obtaining macroscopic physical behavior, but this role is, actually, not disputed here, it is just not given the utmost significance. One certainly recognizes that the possibility to prepare a ``past'' state depends crucially also on how large the system is. Moreover, what is claimed, e.~g., in~\cite{CohenRondoni98}, is that ``the equality between the physical entropy production and the dynamical phase space contraction rates of Hamiltonian thermostatted systems only holds for macroscopic systems''. Restricting the scope of the paper to the investigation of irreversibility only, such arguments do not apply to it.

\section{Conclusions}

We have analyzed the parametrized set of generalized baker maps as simple models of time-reversible evolution, exhibiting signs of irreversibility. The analysis has shown that it is possible to define past, present, and future  states and distribution functions on the definition domain of the map. The question about the origin of the preferred orientation of evolution --- the arrow of time --- can then be answered by stating that it is rooted in contractivity of the map: the latter produces a unique direction of evolution which can be brought into relation with the arrow. This applies equally to the measure-preserving classical baker map.

In an effort to view the map as a model --- albeit a very crude one --- of real systems, and to apply the results of the model to them, one sees that a kind of correspondence can be established between GBM and macroscopic systems, the dynamics of which has an attractor/repellor pair (visible or hidden). The second law of thermodynamics for such systems can then be explained in the same way as the unique direction of evolution in GBM. Namely, it can be regarded as a  consequence of the dynamics which --- if continued to encompass the past --- would lead from the past repellor to the future attractor. The observed properties of the systems then result from this simple symmetry. Macroscopic observation of systems --- the sole viewpoint available to us --- hides this fundamental microscopic property and causes a series of seeming contradictions between the two directions of time. Accepting this standpoint, we can claim that there is no real ``problem'' with the law and that the objections posed against it (including the classical ones of Loschmidt and Zermelo) are simply the results of misunderstanding.

I think that the described model essentially explains the origin of observed (macroscopic) irreversibility in unobservably (microscopically) time-reversible systems. The observed macroscopic irreversibility is actually reversible \emph{in principle}, or the microscopic reversibility is observationally irreversible. The driving force of such behavior is a sensitive dependence of the evolution on initial conditions, which represents the hallmark of chaos.

\begin{acknowledgments}
The final stage in the preparation of the paper has profited much from the discussions with William G.~Hoover and Harald A.~Posch. This work was partly supported by the Scientific Grant Agency (VEGA) of the Slovak Academy of Science and Ministry of Education of the Slovak Republic under the Grant No. 1/0428/03. The support is herewith greatly acknowledged.
\end{acknowledgments}

\bibliography{Kumicak}

\begin{thebibliography}{24}
\expandafter\ifx\csname natexlab\endcsname\relax\def\natexlab#1{#1}\fi
\expandafter\ifx\csname bibnamefont\endcsname\relax
  \def\bibnamefont#1{#1}\fi
\expandafter\ifx\csname bibfnamefont\endcsname\relax
  \def\bibfnamefont#1{#1}\fi
\expandafter\ifx\csname citenamefont\endcsname\relax
  \def\citenamefont#1{#1}\fi
\expandafter\ifx\csname url\endcsname\relax
  \def\url#1{\texttt{#1}}\fi
\expandafter\ifx\csname urlprefix\endcsname\relax\def\urlprefix{URL }\fi
\providecommand{\bibinfo}[2]{#2}
\providecommand{\eprint}[2][]{\url{#2}}

\bibitem[{\citenamefont{Mackey}(1992)}]{Mackey92}
\bibinfo{author}{\bibfnamefont{M.~C.} \bibnamefont{Mackey}},
  \emph{\bibinfo{title}{Time's Arrow---The Origins of Thermodynamic Behaviour}}
  (\bibinfo{publisher}{Springer-Verlag}, \bibinfo{address}{New York},
  \bibinfo{year}{1992}).

\bibitem[{\citenamefont{Spohn}(1991)}]{Spohn91}
\bibinfo{author}{\bibfnamefont{H.}~\bibnamefont{Spohn}},
  \emph{\bibinfo{title}{Large Scale Dynamics of Interacting Particles}}
  (\bibinfo{publisher}{Springer-Verlag}, \bibinfo{address}{Berlin},
  \bibinfo{year}{1991}).

\bibitem[{\citenamefont{Kumi\v{c}\'{a}k and
  Br\"{a}ndas}(1987)}]{KumicakBrandas87a}
\bibinfo{author}{\bibfnamefont{J.}~\bibnamefont{Kumi\v{c}\'{a}k}}
  \bibnamefont{and}
  \bibinfo{author}{\bibfnamefont{E.}~\bibnamefont{Br\"{a}ndas}},
  \bibinfo{journal}{Int. J. Quantum Chem.} \textbf{\bibinfo{volume}{32}},
  \bibinfo{pages}{669} (\bibinfo{year}{1987}).

\bibitem[{\citenamefont{Spohn}(1997)}]{Spohn97}
\bibinfo{author}{\bibfnamefont{H.}~\bibnamefont{Spohn}}, in
  \emph{\bibinfo{booktitle}{Pioneering Ideas for the Physical and Chemical
  Sciences}}, edited by \bibinfo{editor}{\bibnamefont{Fleischhacker}}
  \bibnamefont{and} \bibinfo{editor}{\bibnamefont{Sch\"{o}nfeld}}
  (\bibinfo{publisher}{Plenum Press}, \bibinfo{year}{1997}), pp.
  \bibinfo{pages}{153--157}.

\bibitem[{\citenamefont{de~Hemptinne}(1992)}]{Xavier92}
\bibinfo{author}{\bibfnamefont{X.}~\bibnamefont{de~Hemptinne}},
  \emph{\bibinfo{title}{Non-equilibrium Statistical Thermodynamics applied to
  Fluid Dynamics and Laser Physics}} (\bibinfo{publisher}{World Scientific},
  \bibinfo{address}{Singapore}, \bibinfo{year}{1992}).

\bibitem[{\citenamefont{Kumi\v{c}\'ak and
  de~Hemptinne}(1998)}]{KumicakHemptinne98}
\bibinfo{author}{\bibfnamefont{J.}~\bibnamefont{Kumi\v{c}\'ak}}
  \bibnamefont{and}
  \bibinfo{author}{\bibfnamefont{X.}~\bibnamefont{de~Hemptinne}},
  \bibinfo{journal}{Physica D} \textbf{\bibinfo{volume}{112}},
  \bibinfo{pages}{258} (\bibinfo{year}{1998}).

\bibitem[{\citenamefont{Hoover}(1999)}]{HooverRev}
\bibinfo{author}{\bibfnamefont{W.~G.} \bibnamefont{Hoover}},
  \emph{\bibinfo{title}{Time Reversibility, Computer Simulation, and Chaos}}
  (\bibinfo{publisher}{World Scientific}, \bibinfo{address}{Singapore},
  \bibinfo{year}{1999}).

\bibitem[{\citenamefont{Schuster}(1988)}]{Schuster88}
\bibinfo{author}{\bibfnamefont{H.~G.} \bibnamefont{Schuster}},
  \emph{\bibinfo{title}{Deterministic Chaos: An Introduction}}
  (\bibinfo{publisher}{VCH}, \bibinfo{address}{Weinheim},
  \bibinfo{year}{1988}).

\bibitem[{\citenamefont{Dorfman}(1999)}]{Dorfman}
\bibinfo{author}{\bibfnamefont{J.~R.} \bibnamefont{Dorfman}},
  \emph{\bibinfo{title}{An introduction to chaos in nonequilibrium mechanics}}
  (\bibinfo{publisher}{Cambridge University Press},
  \bibinfo{address}{Cambridge}, \bibinfo{year}{1999}).

\bibitem[{\citenamefont{Kumi\v{c}\'{a}k}(2001)}]{Kumicak01}
\bibinfo{author}{\bibfnamefont{J.}~\bibnamefont{Kumi\v{c}\'{a}k}},
  \bibinfo{journal}{Chaos} \textbf{\bibinfo{volume}{11}}, \bibinfo{pages}{624}
  (\bibinfo{year}{2001}).

\bibitem[{\citenamefont{Ott}(1993)}]{Ott}
\bibinfo{author}{\bibfnamefont{E.}~\bibnamefont{Ott}},
  \emph{\bibinfo{title}{Chaos in Dynamical Systems}}
  (\bibinfo{publisher}{Cambridge University Press},
  \bibinfo{address}{Cambridge}, \bibinfo{year}{1993}).

\bibitem[{\citenamefont{Guckenheimer and Holmes}(1983)}]{Guckenheimer83}
\bibinfo{author}{\bibfnamefont{J.}~\bibnamefont{Guckenheimer}}
  \bibnamefont{and} \bibinfo{author}{\bibfnamefont{P.}~\bibnamefont{Holmes}},
  \emph{\bibinfo{title}{Nonlinear Oscillations, Dynamical Systems, and
  Bifurcations of Vector Fields}} (\bibinfo{publisher}{Springer-Verlag},
  \bibinfo{address}{New York}, \bibinfo{year}{1983}).

\bibitem[{\citenamefont{Hoover and Posch}(1998)}]{Hoover98}
\bibinfo{author}{\bibfnamefont{W.~G.} \bibnamefont{Hoover}} \bibnamefont{and}
  \bibinfo{author}{\bibfnamefont{H.~A.} \bibnamefont{Posch}},
  \bibinfo{journal}{Chaos} \textbf{\bibinfo{volume}{8}}, \bibinfo{pages}{366}
  (\bibinfo{year}{1998}).

\bibitem[{\citenamefont{Falconer}(1990)}]{Falconer90}
\bibinfo{author}{\bibfnamefont{K.}~\bibnamefont{Falconer}},
  \emph{\bibinfo{title}{Fractal Geometry---Mathematical Foundations and
  Applications}} (\bibinfo{publisher}{John Wiley},
  \bibinfo{address}{Chichester}, \bibinfo{year}{1990}).

\bibitem[{\citenamefont{Barnsley}(1988)}]{Barnsley88}
\bibinfo{author}{\bibfnamefont{M.}~\bibnamefont{Barnsley}},
  \emph{\bibinfo{title}{Fractals Everywhere}} (\bibinfo{publisher}{Academic
  Press}, \bibinfo{address}{London}, \bibinfo{year}{1988}).

\bibitem[{\citenamefont{Misra et~al.}(1979)\citenamefont{Misra, Prigogine, and
  Courbage}}]{MPC79}
\bibinfo{author}{\bibfnamefont{B.}~\bibnamefont{Misra}},
  \bibinfo{author}{\bibfnamefont{I.}~\bibnamefont{Prigogine}},
  \bibnamefont{and} \bibinfo{author}{\bibfnamefont{M.}~\bibnamefont{Courbage}},
  \bibinfo{journal}{Physica} \textbf{\bibinfo{volume}{A98}}, \bibinfo{pages}{1}
  (\bibinfo{year}{1979}).

\bibitem[{\citenamefont{Antoniou and Misra}(1991)}]{AntoniouMisra91}
\bibinfo{author}{\bibfnamefont{I.~E.} \bibnamefont{Antoniou}} \bibnamefont{and}
  \bibinfo{author}{\bibfnamefont{B.}~\bibnamefont{Misra}}, \bibinfo{journal}{J.
  Phys. A: Math. General} \textbf{\bibinfo{volume}{24}}, \bibinfo{pages}{2723}
  (\bibinfo{year}{1991}).

\bibitem[{\citenamefont{Bricmont}(1996)}]{Bricmont96}
\bibinfo{author}{\bibfnamefont{J.}~\bibnamefont{Bricmont}}, in
  \emph{\bibinfo{booktitle}{Ann. New York Acad. Sci., Vol. 775: ``The Flight
  from Science and Reason''}}, edited by \bibinfo{editor}{\bibfnamefont{P.~R.}
  \bibnamefont{Gross}},
  \bibinfo{editor}{\bibfnamefont{N.}~\bibnamefont{Levitt}}, \bibnamefont{and}
  \bibinfo{editor}{\bibfnamefont{M.~W.} \bibnamefont{Lewis}}
  (\bibinfo{publisher}{The New York Academy of Science}, \bibinfo{address}{New
  York}, \bibinfo{year}{1996}), pp. \bibinfo{pages}{131--175}.

\bibitem[{\citenamefont{Kumi\v{c}\'{a}k}(1977)}]{Kumicak77}
\bibinfo{author}{\bibfnamefont{J.}~\bibnamefont{Kumi\v{c}\'{a}k}},
  \bibinfo{journal}{Czech. J.Phys.} \textbf{\bibinfo{volume}{B27}},
  \bibinfo{pages}{967} (\bibinfo{year}{1977}).

\bibitem[{\citenamefont{Cohen and Rondoni}(1998)}]{CohenRondoni98}
\bibinfo{author}{\bibfnamefont{E.~G.~D.} \bibnamefont{Cohen}} \bibnamefont{and}
  \bibinfo{author}{\bibfnamefont{L.}~\bibnamefont{Rondoni}},
  \bibinfo{journal}{Chaos} \textbf{\bibinfo{volume}{8}}, \bibinfo{pages}{357}
  (\bibinfo{year}{1998}).

\bibitem[{\citenamefont{Stoddard and Ford}(1973)}]{Stoddard73}
\bibinfo{author}{\bibfnamefont{S.~D.} \bibnamefont{Stoddard}} \bibnamefont{and}
  \bibinfo{author}{\bibfnamefont{J.}~\bibnamefont{Ford}},
  \bibinfo{journal}{Phys. Rev. A} \textbf{\bibinfo{volume}{8}},
  \bibinfo{pages}{1504–} (\bibinfo{year}{1973}).

\bibitem[{\citenamefont{Dellago et~al.}(1996)\citenamefont{Dellago, Posch, and
  Hoover}}]{Dellago96}
\bibinfo{author}{\bibfnamefont{C.}~\bibnamefont{Dellago}},
  \bibinfo{author}{\bibfnamefont{H.~A.} \bibnamefont{Posch}}, \bibnamefont{and}
  \bibinfo{author}{\bibfnamefont{W.~G.} \bibnamefont{Hoover}},
  \bibinfo{journal}{Phys.~Rev.~E} \textbf{\bibinfo{volume}{53}},
  \bibinfo{pages}{1485} (\bibinfo{year}{1996}).

\bibitem[{\citenamefont{Dobrushin et~al.}(2000)\citenamefont{Dobrushin, Sinai,
  and Sukhov}}]{Sinai00}
\bibinfo{author}{\bibfnamefont{R.~L.} \bibnamefont{Dobrushin}},
  \bibinfo{author}{\bibfnamefont{Y.~G.} \bibnamefont{Sinai}}, \bibnamefont{and}
  \bibinfo{author}{\bibfnamefont{Y.~M.} \bibnamefont{Sukhov}}, in
  \emph{\bibinfo{booktitle}{Dynamical Systems, Ergodic Theory and
  Applications}}, edited by \bibinfo{editor}{\bibfnamefont{Y.~G.}
  \bibnamefont{Sinai}} (\bibinfo{publisher}{Springer}, \bibinfo{year}{2000}),
  pp. \bibinfo{pages}{384--429}.

\bibitem[{\citenamefont{Hahn}(1953)}]{Hahn53}
\bibinfo{author}{\bibfnamefont{E.~L.} \bibnamefont{Hahn}},
  \bibinfo{journal}{Physics Today} \textbf{\bibinfo{volume}{6}},
  \bibinfo{pages}{4} (\bibinfo{year}{1953}).

\end{thebibliography}

\end{document}